\documentclass[12pt]{iopart}

\RequirePackage{mathtools}

\usepackage{iopams}
\usepackage{mathrsfs}
\usepackage{amsfonts}
\usepackage{latexsym}
\usepackage{amsthm}
\usepackage{amsmath}
\usepackage{graphicx}
\usepackage{braket}
\usepackage{xcolor}
\usepackage{subfigure}
\usepackage{CJK}\usepackage{enumitem}
\usepackage{physics}
\usepackage{scrextend}
\usepackage{multirow}

\definecolor{wwblue}{RGB}{7,81,159}

\makeatletter
\newcommand\footnoteref[1]{\protected@xdef\@thefnmark{\ref{#1}}\@footnotemark}
\makeatother

\begin{document}
\begin{CJK*}{UTF8}{gbsn}
\title{Inter-species entanglement of Bose-Bose mixtures trapped in optical lattices}

\author{Wei Wang (王巍)$^1$, Vittorio Penna$^2$, Barbara Capogrosso-Sansone$^3$}

\address{$^1$ Homer L. Dodge Department of Physics and Astronomy, The University of Oklahoma, Norman, Oklahoma 73019, USA}
\address{$^2$ Dipartimento di Scienza Applicata e Tecnologia, Politecnico di Torino, Corso Duca degli Abruzzi 24, I-10129 Torino, Italy and CNISM, u.d.r., Politecnico di Torino, Corso Duca degli Abruzzi 24, I-10129 Torino, Italy}
\address{$^3$ Department of Physics, Clark University, Worcester, Massachusetts 01610, USA}

\ead{wei@ou.edu}

\begin{abstract}
In the present work we discuss inter-species entanglement in Bose-Bose mixtures trapped in optical lattices. This work is motivated by the observation that, in the presence of a second component, the Mott-insulator lobe shifts {\em{differently}} on the hole- and particle-side with respect to the Mott lobe of the single species system~\cite{Barbara2,Barbara1}. We use perturbation theory, formulated in a Hilbert space decomposed by means of lattice symmetries, in order to show that the nonuniform shift of the Mott lobe is a consequence of an inter-species entanglement which differs in the lowest excited states to remove and add a particle. Our results indicate that inter-species entanglement in mixtures can provide a new perspective in understanding quantum phase transitions. To validate our approach, we compare our results from perturbation theory with quantum Monte Carlo simulations.
%
\end{abstract}


\maketitle

\end{CJK*}

\section{Introduction}

In the past decade, quantum phase transitions of Bose-Bose, Bose-Fermi and Fermi-Fermi mixtures in optical lattices have attracted considerable attention both experimentally and theoretically~\cite{quantummagneticphase2,bosonbosonexperiment1,bosonbosonexperiment2,bosonbosonexperiment3,Iskin1,Kuklov1,Kuklov2,Isacsson1,Pai1,Ozaki1,Nakano1,FermiBose1,FermiBose2,FermiBose3,FermiBose4,FermiBose5,FermiBose6,fermiboseexperiment1,fermiboseexperiment2,FermiFermi1,FermiFermi2,fermifermiexperiment1}. Nonetheless, the physics of mixtures is still not completely understood. This is because the coupling between the various components introduces extra degrees of freedom which result in a wealth of exotic and novel phases. Indeed, mixtures feature quantum phase transitions otherwise absent in single species systems. Moreover, the inter-species coupling introduces non-trivial correlations between components which may result in significant inter-species entanglement. Inter-species entanglement can offer a different perspective in understanding quantum phase transitions in multicomponent systems. Hence,
by applying concepts from quantum computation to many-body physics~\cite{entanglementbosehubbard1,entanglementbosehubbard2,entanglementbosehubbard3,entanglementbosehubbard4,enbh1}, one can gain further insight into the physics of mixtures.

Recent theoretical and experimental results report on the modification of the Mott lobe and, in general, of the Mott-insulator to superfluid transition~\cite{Fisher1,Jaksch1,Greiner1} in the presence of a second component~\cite{Barbara2,Barbara1,FermiBose1,FermiBose2,fermiboseexperiment2}. In particular, it has been theoretically shown and experimentally observed that the influence of a second bosonic or fermionic species on the insulating phases of the other bosonic component can be controlled by the strength of the inter-species coupling and by the density of the second component. Intuitively, the strength of the inter-species coupling can be viewed as an indicator of how entangled the two species are. In particular, in the limit of negligible coupling between the two components, the same physics as for a single component system would be observed and inter-species entanglement would be absent. Thus, naturally, one can expect that the modifications on the Mott insulator lobe in the presence of a second component are a manifestation of inter-species entanglement.

In this article, we consider Bose-Bose mixtures trapped in optical lattices. Our main goal is to provide a {{\em qualitative}} understanding of how inter-species entanglement relates to the observations made in~\cite{Barbara2,Barbara1} where the authors report a visible shift of the Mott lobe boundary on the hole-side (i.e. when the superfluid is reached by doping with holes) and an almost negligible shift on the particle-side of the lobe (i.e.  when the superfluid is reached by doping with particles).

This paper is organized as follows. In Section~\ref{sec2} we introduce the two-component Bose-Hubbard model. In Section~\ref{sec3} we provide intuitive understanding of how inter-species entanglement affects the Mott-insulator to superfluid phase transition in the binary mixture. In Section~\ref{sec4} we introduce the theoretical framework on which perturbation theory is based and discuss the symmetries of the system. In Section \ref{sec5} we discuss inter-species entanglement of the two-component Bose-Hubbard model defined in a reduced and decomposed Hilbert space.  In Section~\ref{sec6} we present numerical results based on perturbative calculation and we compare them with exact quantum Monte Carlo results. Finally, we conclude in Section~\ref{sec8}.

\section{The model}
\label{sec2}
We consider a bosonic binary mixture trapped in an optical lattice as described by the two-component Bose-Hubbard model
\begin{eqnarray}
\label{Eq0}
H=\frac{{U^a}}{2}\sum_{i=1}^ M n^a_{i}(n^a_{i}-1)
+\frac{{U^b}}{2}\sum_{i=1}^ M n^b_{i}(n^b_{i}-1)+{U^{ab}} \sum_{i=1}^M n^a_i n^b_{i} \nonumber\\
\quad\quad-{T^a}\sum_{(i,j)}\, a^{\dagger}_i a_j 
-{T^b}\sum_{(i,j)}\, b^{\dagger}_i b_j 
-\mu^{a}N^{a}-\mu^{b}N^{b},
\end{eqnarray}
where $M$ is the number of lattice sites, $a$ stands for the first species (species-a or component-a) while $b$ stands for the second species (species-b or component-b), $a_i^\dagger(b_i^\dagger)$ and $a_i(b_i)$ are creation and annihilation operators of species-a(b) at site $i$, $n_i^a=a^\dagger_i a_i$, $n_i^b=b^\dagger_i b_i$, are the particle number operators at site $i$ and $N^a=\sum_{i}n_i^a$, $N^b=\sum_{i}n_i^b$. $U^a$ and $U^b$ are the onsite intra-species interactions for component-a and -b respectively, $U^{ab}$ is the inter-species onsite interaction, $T^a$ and $T^b$ are the hopping amplitudes, and $\mu^a$ and $\mu^b$ are the chemical potentials which set the total number of particles. In the following we consider species-a to be the majority species, whose Mott lobe boundary is affected by the presence of species-b, the minority component.

\section{Characterization of the shift of the lobe boundary in terms of mutual information}
\label{sec3}
In this Section we discuss previous observations of a shift of the Mott-insulator (MI) lobe boundary in the presence of a second bosonic minority component, and consider these observations in the context of inter-species entanglement and mutual information of states. 

In References~\cite{Barbara2,Barbara1} it was found that, in the presence of species-b as a minority component, the boundary of MI lobe of species-a is affected differently on the hole- and particle-side compared to the MI lobe in the absence of species-b.
Overall, the shift of the MI lobe is always more prominent on the hole-side of the boundary. The magnitude and the modality of the shift depend on the interplay between kinetic and potential energies, and the density of species-b. 
For example, at fixed $U_{ab}$ and for a given density of species-b,  the first MI lobe of species-a always possesses a visible shift on the hole-side of the boundary with respect to the lobe in the single-species case, while the shift on the particle-side of the boundary is either absent or considerably less pronounced, depending on the density of species-b~\cite{Barbara2,Barbara1}. In particular, the particle-side of the boundary shifts only for sufficiently large fillings of component-b. It is worth noting that the lobe boundaries can either be explored by fixing $U^a$ while varying $T^a$ or vice versa. In the former case, the boundary shift on the hole-side is already observed in the limit $T^a/U^a\rightarrow 0$ and it gets progressively smaller as the hopping is increased and quantum fluctuations become more prominent. On the other hand, when the boundary is explored by fixing $T^a$ and varying $U^a$, the boundary shift is inexistent in the limit $T^a/U^a\rightarrow 0$ for which, the onsite intra-species interaction $U^a$ becomes the dominant energy scale and the interaction between the two components can be neglected. The shift becomes progressively larger upon increasing $T^a/U^a$. 
In general, although the modality of the boundary shift depend on the specific choice of model parameters, overall, a larger inter-species interaction leads to a greater shift of the lobe.

In the following, we focus on exploring the lobe boundary by fixing $U^a$ while varying $T^a$. As discussed above, the shift is already visible in the limit $T^a/U^a\rightarrow 0$ which implies that one can perform a perturbative study of the phase boundary shift by treating the hopping term as the perturbation. In order to gain a qualitative understanding of the role played by species-b, let's first consider the limit $T^a\rightarrow 0$, $T^b\rightarrow0$ and consider $N_b=1$. 
Throughout this paper, we assume all onsite interactions to be repulsive and inter-species interactions are chosen to avoid phase segregation~\cite{Kuklov1}. We consider a square lattice with periodic boundary conditions.

\begin{figure}
\centering
\begin{tabular}{r l}
\multirow{ 4}{*} {\subfigure[\label{lobe}]{\includegraphics[width=0.53\textwidth]{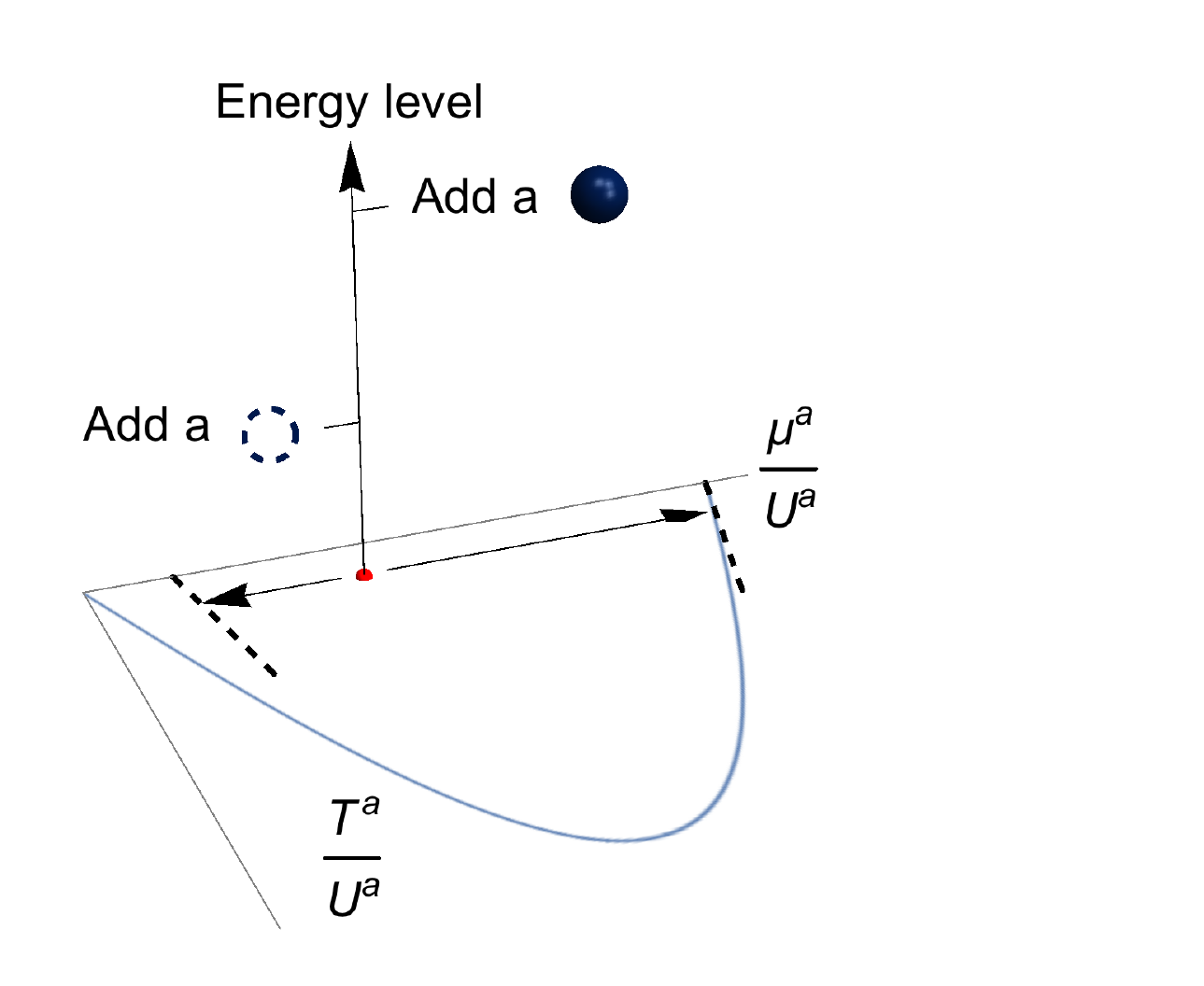}}} &\\
&\subfigure[\label{holeside}]{\includegraphics[width=0.19\textwidth]{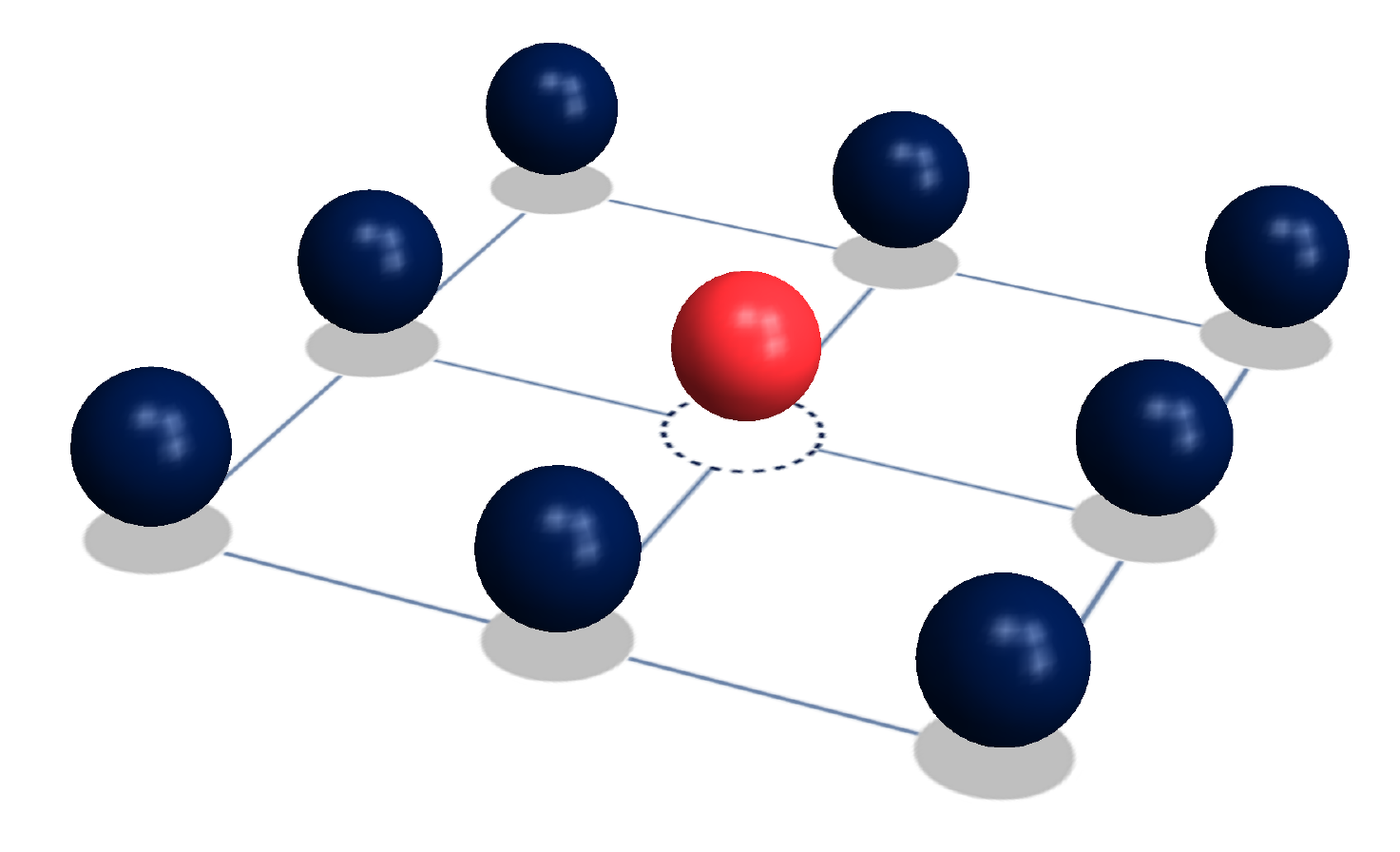}}\\
&\subfigure[\label{particleside}]{\includegraphics[width=0.19\textwidth]{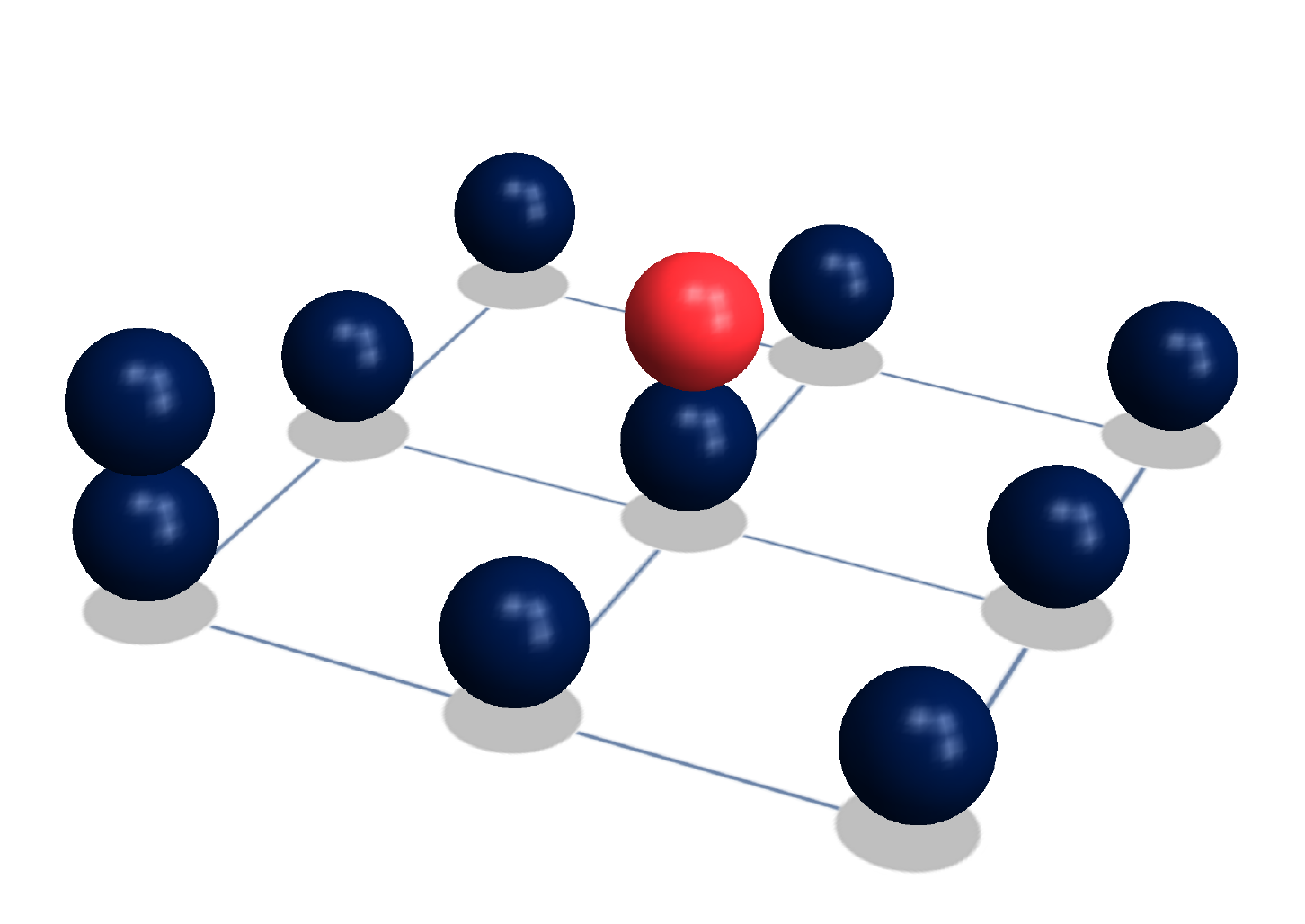}}\\
&\subfigure[\label{ground}]{\includegraphics[width=0.19\textwidth]{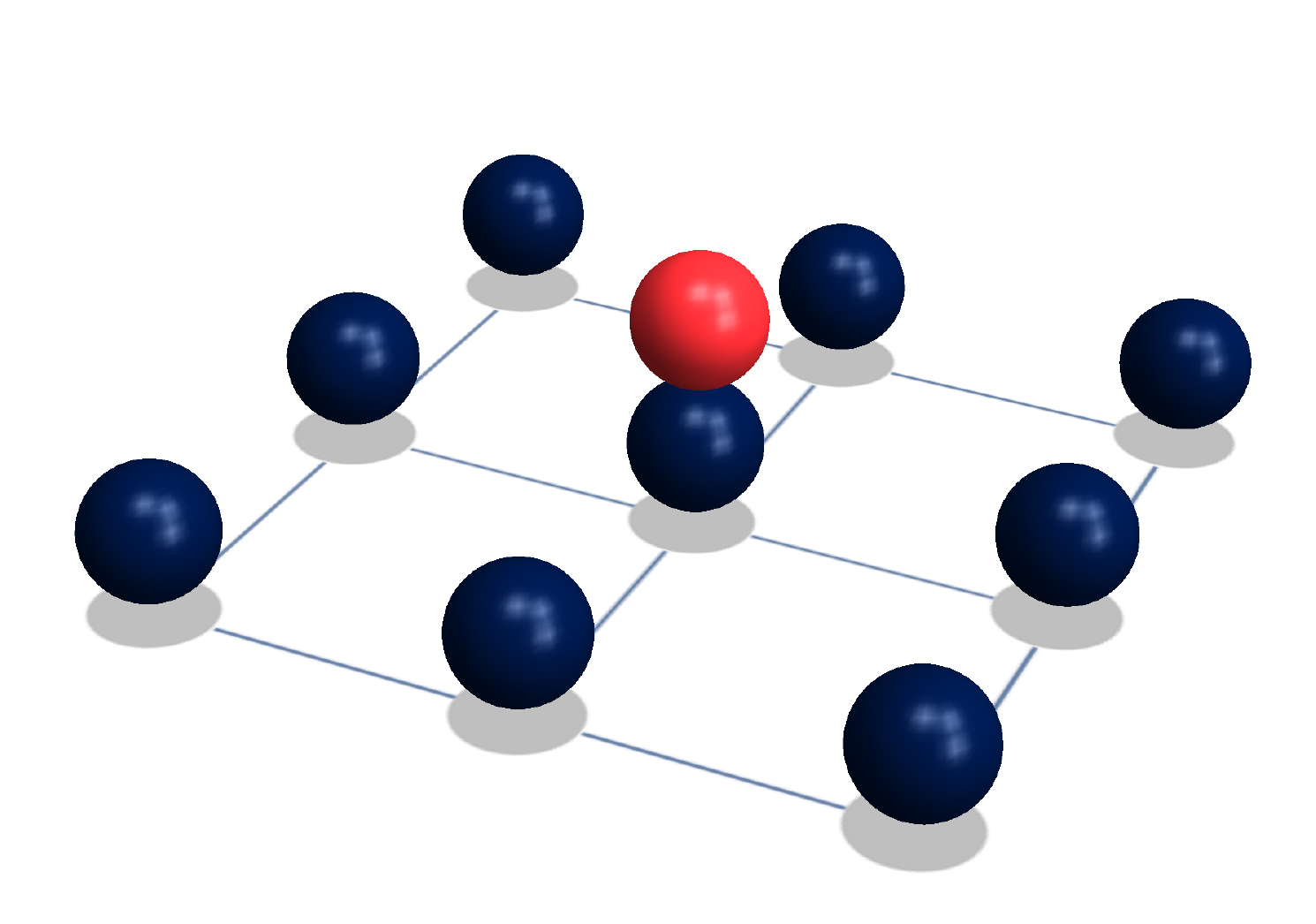}}
\end{tabular}

\caption{\label{FigOne}a) Sketch of the Mott insulator lobe of the single component (solid blue line). Sketch of the shifted lobe in the presence of a second component (dashed black line). The red spot inside the lobe represent at fixed value of chemical potential with arrows indicating the gap the add a hole or an extra particle. The excited energy corresponding to the addition of a hole (dotted circle) and an extra particle (solid circle) are sketched on the vertical axis. b) Representative Fock state when a hole is added to the Mott insulator. The hole of component-a and particle-b occupy the same lattice site in order to minimize the energy. c) Representative Fock state when an extra particle is added to the Mott insulator. The extra particle of component-a occupies a site different than the one occupied by particle-b. d) Representative Fock state in the Mott insulator of component-a in the presence of particle-b. 
}
\end{figure}

Let us start by reviewing the determination of the boundaries of the MI lobe away from the tip. In the grand canonical ensemble, the MI lobe of species-a is a result of the energy gap between the ground state and the elementary excited state which corresponds to adding a hole or a particle~\cite{Fisher1} of species-a to the mixture. Denoting the ground state energy by $E$ and the lowest excited energy to add a hole (particle) by $E_h$ ($E_p$), one has $E_h>E$ ($E_p>E$) inside the MI lobe, and $E_h=E$ ($E_p=E$) at the hole-side (particle-side) boundary. Hence, one can find the position of the boundary by setting the lowest excited energy equal to the ground-state energy. It is therefore easy to understand the boundary shift in terms of a change of the energy gap induced by the presence of the second component. For a given chemical potential inside the MI lobe, the hole-side (particle-side) gap is given by the distance between the chosen chemical potential and the hole (particle) boundary as shown in Figure~\ref{lobe}. Without the species-b boson and in the zero hopping limit, we have
\begin{equation*}
E=-\mu^aM,\quad E_h=-\mu^a(M-1),\quad E_p=U^a-\mu^a(M+1)
\end{equation*}
and the gaps $E_h-E=\mu^a$, $E_p-E=U^a-\mu^a$. In the presence of the species-b boson, the repulsive inter-species interaction ``helps'' removing a species-a boson from the lattice in order to create a hole, so that the presence of species-b lowers the excitation energy needed to add a hole, thus shrinking the lobe on the hole-side as shown in Figure~\ref{lobe}. This can be easily understood in the limit of zero hopping where, in the presence of a hole of species-a, particle-b occupies the same site as the hole-a in order to minimize the inter-species interaction (see Figure~\ref{holeside}). On the other hand, the repulsive interaction between the two species has no influence on the energy needed to add a particle since, in order to minimize the inter-species interaction, the added species-a boson will not occupy the same site as the species-b boson as shown in Figure~\ref{particleside}. Hence, the excitation energy to add a particle remains unchanged and the lobe shift is absent. Correspondingly, in the presence of particle-b we have 
\begin{equation*}
E=U^{ab}-\mu^aM-\mu^b,\quad E_h=-\mu^a(M-1)-\mu^b,\quad E_p=U^a+U^{ab}-\mu^a(M+1)-\mu^b
\end{equation*}
and the gaps $E_h-E=\mu^a-U^{ab}$, $E_p-E=U^a-\mu^a$. Obviously, the energy gap to add a hole is lowered by $U^{ab}$ while the energy gap to add a particle stays the same.

Let us now turn to the discussion of how inter-species entanglement is related to the non-uniform MI lobe shift we just described. In the system considered, the inter-species entanglement arises from a tensor-product form $\mathscr{H}^a\bigotimes\mathscr{H}^b$ of the Hilbert space of the system, where $\mathscr{H}^a$ and $\mathscr{H}^b$ are the Fock spaces corresponding to each single-species system.
In the limit of zero hopping~\footnote{Here, we consider the limit of zero hopping $T^a\rightarrow 0$, $T^b\rightarrow 0$  instead of exactly zero hopping $T^a=T^b=0$ in order to exploit the results discussed in~\cite{wei1} which state that, at non-zero hopping, the ground states of the two-component Bose-Hubbard model corresponding to fixed particle numbers are always {\em{unique}}. This means that, at finite hopping, the system is described by a pure state rather than a density operator.}, the ground state of the MI and the first excited states to add a hole or a particle are expanded in terms of Fock states each with equal coefficient (weight). Moreover the Fock states entering the expansion are characterized by similar particle distribution on the lattice~\cite{wei1} as we explain in detail below.

In the Fock states spanning the MI ground state, species-a bosons are uniformly located in the lattice while the single species-b boson can be located anywhere (an example of this type of configurations is shown in Figure~\ref{ground}). It is clear that the location of species-a is {\em{irrelevant}} to the location of species-b. In terms of {\em{mutual information}}, which, in this context, represents the ability of determining e.g. the position of particle-b from the position of particles-a, we can conclude that in the MI ground state there is no mutual information between the two species. For pure states, mutual information can also be seen as a measure of the entanglement~\cite{entanglementbook1}, hence, we expect the MI ground state to be  non-entangled.

In the Fock states spanning the first excited state to add a hole (in the following we refer to it by hole-side excited state), the single species-b boson has to be located on the same site where the hole-a is located in order to minimize the inter-species interaction. This is shown in Figure~\ref{holeside}. In this case the location of boson-b {\em{uniquely}} determines the location of hole-a. Therefore, we anticipate this state to have maximal mutual information and thus this excited state to be maximally entangled.

In the Fock states spanning the first excited state to add a particle (in the following we refer to it by particle-side excited state), the inter-species interaction is minimized with boson-b located everywhere in the optical lattice except for the site where the extra particle-a is located. An example of this type of configuration is shown in Figure~\ref{particleside}. In this case, the location of particles-a determines the location of boson-b only partially. More specifically, the location of particles-a only specifies the site where boson-b will {\em{not}} be located. Therefore, we anticipate a `small' amount of mutual information between the two species and this state to be `slightly' entangled.  This statement will be quantified in the following, by calculating entanglement entropy as a measure of the entanglement  in the system~\cite{entanglementbook1}. 

The above discussion can be summarized as follows: In the limit of zero hopping, the hole-side boundary has a visible shift corresponding to a maximally entangled hole-side excited state; The particle-side boundary does not shift, corresponding to a not-considerably-entangled particle-side excited state. Indeed, as we will show below by means of a perturbative calculation, the degree of entanglement between the two components is closely related to the extent of the shift of the lobe. Indeed, these observations suggest that the inter-species entanglement differs in the ground state, particle-side excited state and hold-side excited states. This difference leads to different shifts in the energy gaps to add a hole or a particle, hence resulting in a nonuniform shift of the lobe on the two sides. Therefore, it is expected that inter-species entanglement plays a role  in quantum phases transitions of mixtures, specifically, in this case, in the Mott-insulator to superfluid transition at non-fixed number of particles. 

Before introducing the framework of the perturbation theory, it is important to note that for systems described by the Bose-Hubbard model in the grand-canonical ensemble, the first excited states corresponding to adding a hole or a particle are identical to ground states in the canonical ensemble with one less or one extra particle with respect to the ground state of the MI phase. Therefore, it is sufficient to study generic properties of the inter-species entanglement in ground states corresponding to different (fixed) particle numbers. 
With this in mind, in the following we introduce a framework capable of studying inter-species entanglement in ground states corresponding to fixed particle numbers. First, we reduce the Hilbert space of model~\ref{Eq0} utilizing symmetries of the Hamiltonian in order to remove irrelevant degrees of freedom and facilitate numerical study. Second, we decompose the Hilbert space in terms of symmetry and degrees of freedom and give a criterion for inter-species entanglement of ground states.

In our framework, we emphasize the structure of the ground state in relation to the decomposition of Hilbert spaces. We will show that  this decomposition provides simplification and convenience for the numerical implementation of the perturbation theory. Although in the following we consider a square lattice with periodic boundary conditions, our approach can be applied to any type of finite lattice. Moreover, while we focus on repulsive inter-species interaction, the method holds for attractive interaction as well.

\section{Symmetries of the ground state}
\label{sec4}

In this Section we discuss the symmetry of the ground state and utilize it in order 
to reduce the Hilbert space.

Consider the ground state $\Psi$ of the Hamiltonian $H$ given by Equation~\ref{Eq0} 
at fixed particle numbers $N^{a}$ and $ N^{b}$. $\Psi$ lives in the tensor product space 
$\mathscr{H}^a\bigotimes\mathscr{H}^b$ with a Fock basis 
$\{ |{\bf n } \rangle \otimes| {\bf m }\rangle \}$. The spaces $\mathscr{H}^a$ and 
$\mathscr{H}^b$ are finite-dimensional Hilbert spaces of the single-component 
Bose-Hubbard model at fixed particle numbers $ N^{a}$ and $ N^{b}$ respectively. 
They are spanned by Fock bases $\{|{\bf n } \rangle \}$ and  $\{|{\bf m } \rangle \}$, 
determined by the lowest band Wannier functions~\cite{Fisher1,Jaksch1}.

The two-component Bose-Hubbard model is defined on a lattice with a finite number of sites. 
To describe the nearest-neighbor tunneling, the lattice is endowed a graph structure, which 
is specified by bonds or edges~\cite{connected}, i.e. nearest-neighbor pairs of lattice sites. 
Lattice symmetries are described by graph automorphisms. A graph automorphism $g$ is a 
one-to-one mapping on the lattice such that $\{g(i),g(j)\}$ (where $i$ and $j$ are sites) 
is a bond if and only if $\{i,j\}$ is a bond~\cite{connected}. Obviously, all graph 
automorphisms of a finite lattice form a finite group under the function composition 
$(g\circ g')(i)=g(g'(i))$. We denote this group by $G$. In a square lattice 
with periodic boundary condition, $G$ is the group generated by all translations, 
rotations and reflections which leave the lattice unchanged.

\begin{figure}
\centering
\subfigure[\label{fock2}]{\includegraphics[width=0.19\textwidth]{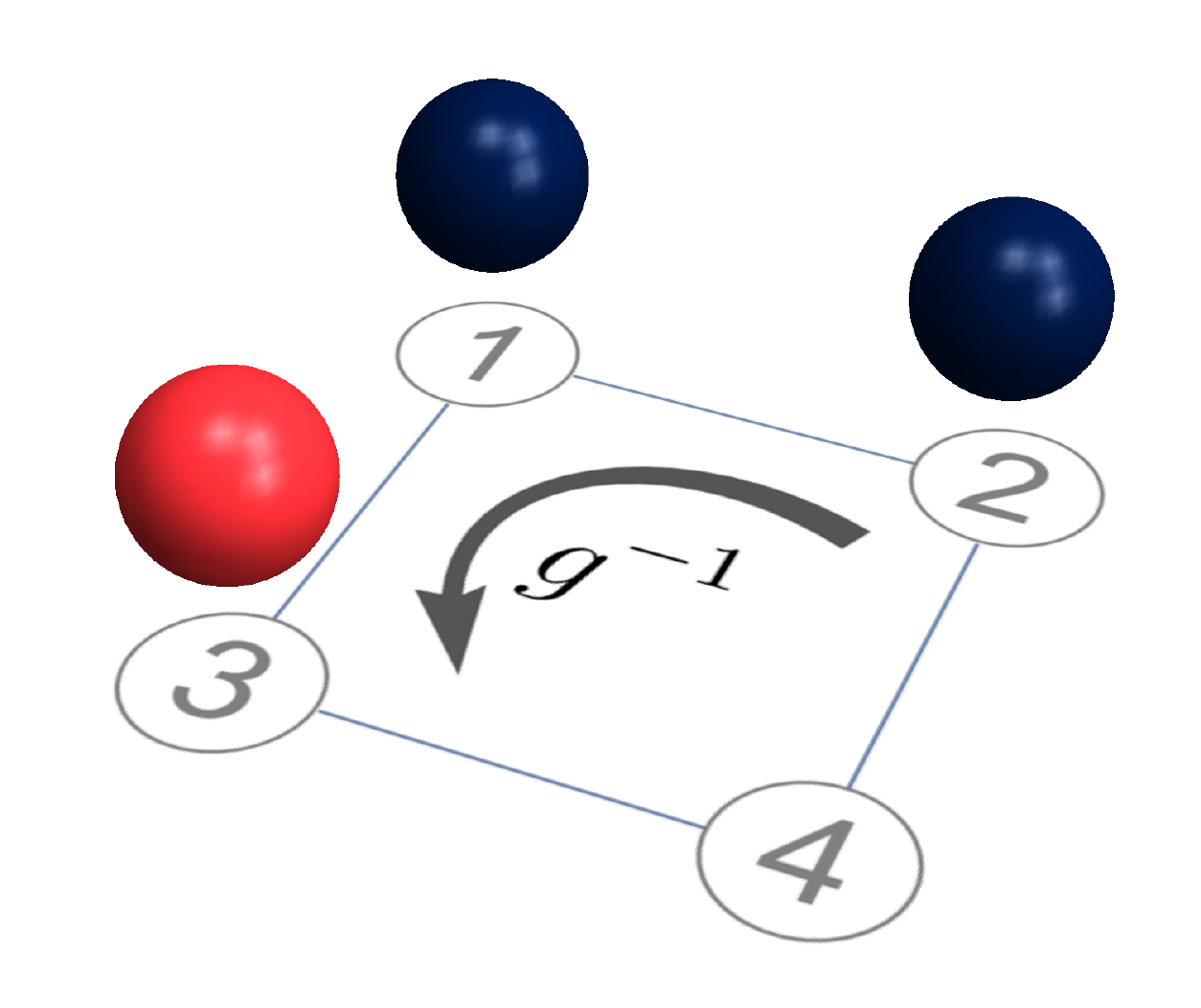}}\quad\quad
\subfigure[\label{fock3}]{\includegraphics[width=0.19\textwidth]{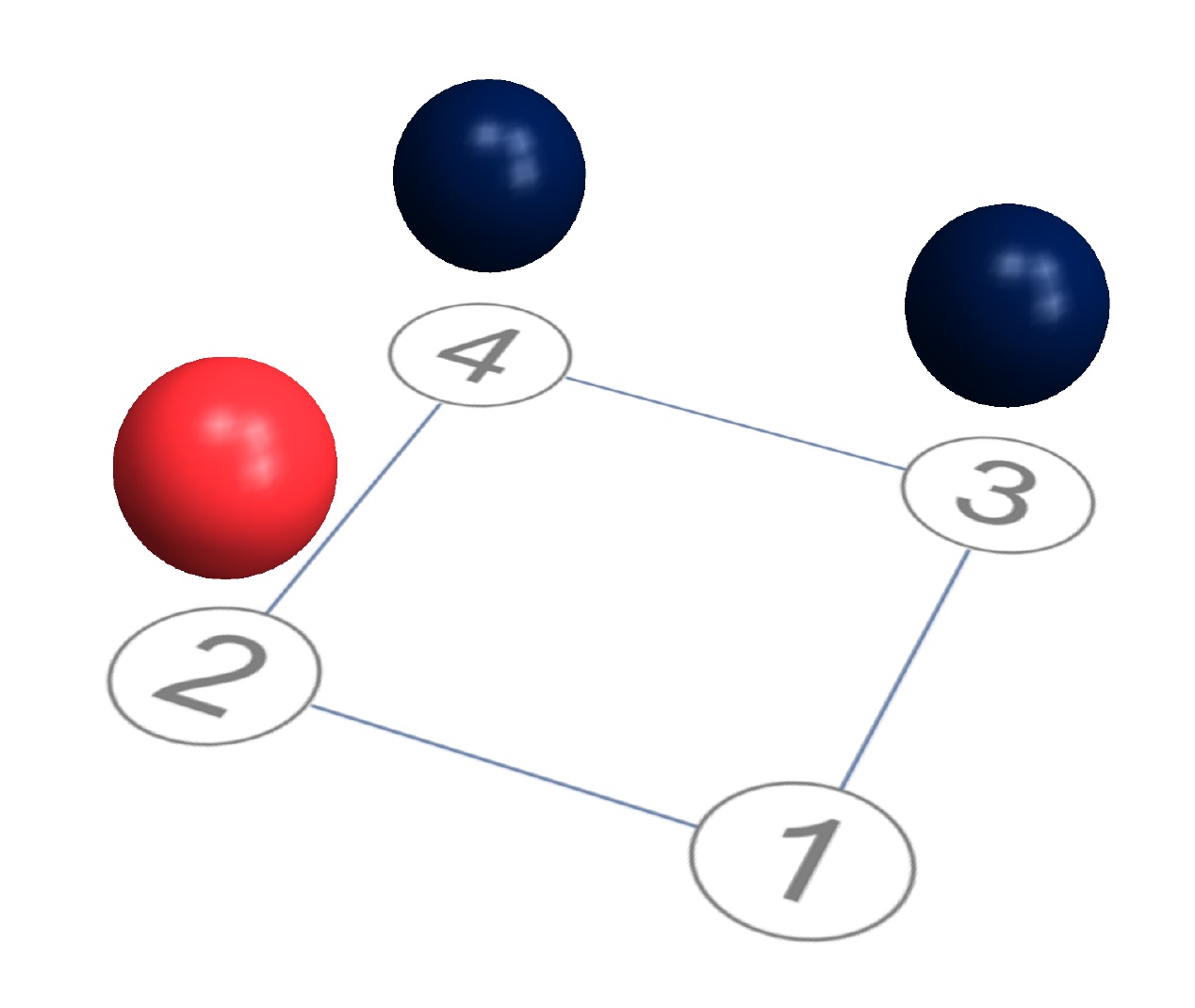}}\quad\quad
\subfigure[\label{fock4}]{\includegraphics[width=0.19\textwidth]{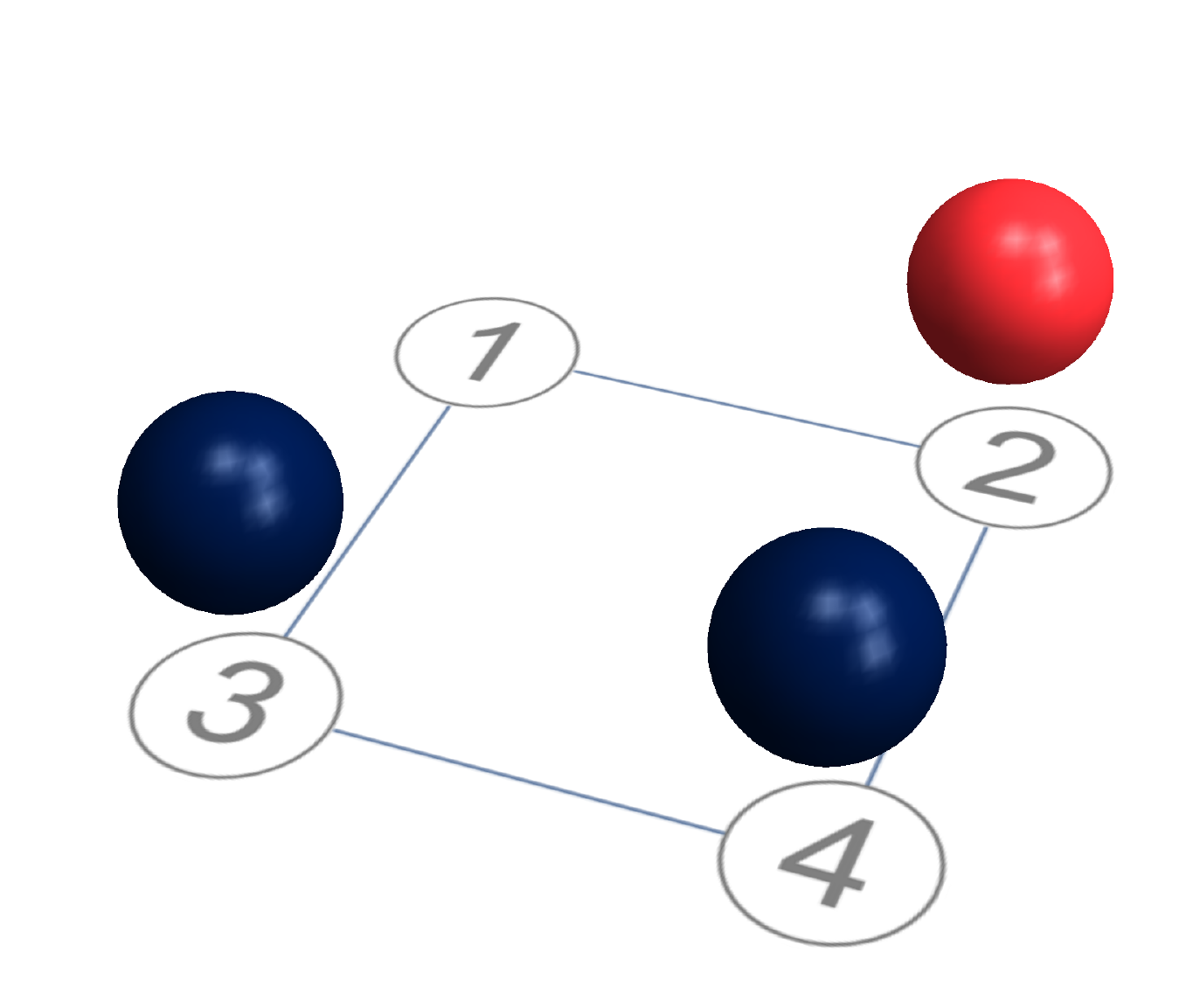}}

\caption{\label{FigTwo}a) and c) 
Sketch of $|{\bf n }, {\bf m }\rangle$ and $C_g |{\bf n }, {\bf m }\rangle$ 
on a $2\times2$ square lattice. Here $g$ represents a $180^\circ$ clockwise rotation. 
b) Intermediate step where, according to the definition of $C_g$, 
the positions of bosons are fixed while $g^{-1}$ operates on the lattice.}
\end{figure}

The group $G$ has three unitary representations 
$\mathcal{\pi}^{ab}$, $\mathcal{\pi}^{a}$ and $\mathcal{\pi}^{b}$ 
in the Hilbert spaces $\mathscr H^{a}\bigotimes\mathscr{H}^{b}$, $\mathscr H^{a}$ and $\mathscr H^{b}$, 
respectively~\footnote{A unitary representation $\mathcal{\pi}$ of the group 
$G$ in the space $\mathscr{H}$ is a group homomorphism mapping each $g$ into
a unitary operator $\mathcal{\pi}(g)$ on $\mathscr{H}$, so that 
$\mathcal{\pi}(g_1)\mathcal{\pi}(g_2) = \mathcal{\pi}(g_1\circ g_2)$.}. 
In general, a representation $\mathcal{\pi}$ is defined by mapping any element $g\in G$ into 
a unitary operator $\mathcal{\pi}(g)$ whose action 
is defined on the relevant Hilbert spaces. 
Here, we use the simplified notation $C_g = \mathcal{\pi}^{ab}(g)$, $A_g =\mathcal{\pi}^{a}(g)$ 
and $B_g = \mathcal{\pi}^{b}(g)$. The representations are naturally defined in terms of Fock bases. 
Specifically, given a Fock state 
$|{\bf n }, {\bf m }\rangle = |{\bf n } \rangle \otimes| {\bf m }\rangle 
= | n_1, n_2 , ... n_i, ...\rangle \otimes| m_1, m_2 , ... m_i, ...\rangle$
(where $n_i$ and $m_i$ are occupation numbers of species-a and species-b bosons on site $i$) in
$\mathscr{H}^{a}\bigotimes\mathscr{H}^{b}$, then 
$ C_g(|{\bf n } \rangle \otimes| {\bf m }\rangle )$ 
is also a Fock state corresponding to 
$$
| {n_{g^{-1}(1)},n_{g^{-1}(2)},\cdots n_{g^{-1}(i)} \cdots} \rangle
\otimes | {m_{g^{-1}(1)},m_{g^{-1}(2)},\cdots m_{g^{-1}(i)}\cdots} \rangle
\; ,
$$ 
where the action of $g$ induces a site reshuffling.
An example of $C_g$ on $|{\bf n }, {\bf m }\rangle$ is illustrated in 
Figure~\ref{FigTwo}. Here, $g$ represents a $180^\circ$ clockwise rotation of a $2\times2$ 
square lattice. Figure~\ref{fock2} displays the initial state 
$|{\bf n} , {\bf m}\rangle$, while Figure~\ref{fock4} displays the final state 
$ C_g |{\bf n }, {\bf m }\rangle$.
%
One can understand the action of $C_g$ from the intermediate step, Figure~\ref{fock3}, where, 
according to the definition of $C_g$, the positions of bosons are fixed while $g^{-1}$ operates on the lattice. In this example $g^{-1}$ is a $180^\circ$ counterclockwise rotation. Hence, the action of 
$C_g$ in going directly from Figure~\ref{fock2} to Figure~\ref{fock4}, can be viewed as the operation 
$g$ on the position of bosons. Similarly, we can define $A$ and $B$, and, according to their 
definition, we have $A_g \otimes B_g= C_g$ for all $g\in G$.

Both, the Hamiltonian $H$ and the ground state $ |\Psi \rangle$ are invariant under the action of 
any graph automorphism $g$, i.e. they are symmetric under the group $G$. The 
invariance  of $H$ under the action of $G$ can be easily seen by observing that 
$[H, C_g]=0$ for any $g\in G$. On the other hand, the invariance of $|\Psi \rangle$ under 
$G$, i.e. $C_g  |\Psi \rangle= |\Psi \rangle$ for any $g\in G$, can be seen
from the following arguments. The ground state energy $E$ is nondegenerate and the expansion 
coefficients of $ |\Psi \rangle$ in the Fock basis are all positive, i.e. 
${\langle {\bf m}, {\bf n} |\Psi \rangle }>0$ for all $|{\bf n }, {\bf m }\rangle$~\cite{wei1}. 
Thus, because any $C_g$ commute with $H$, due to the nondegeneracy of ${E}$, 
$C_g |\Psi \rangle= {c} |\Psi \rangle$, where $c$ is a constant. Here, $c$ must be positive, 
because the matrix elements of $C_g$ in the Fock basis are all real and nonnegative, and 
${\langle {\bf m}, {\bf n}| \Psi \rangle }>0$. 
Moreover, $c$ must be $1$, since the unitary operator 
$C_g$ preserves the norm. So we conclude $C_g |\Psi \rangle= |\Psi \rangle $.

It is also important to notice that from the definition of $C$ which leaves unchanged the 
relative position of bosons, any $|{\bf n }, {\bf m }\rangle$ and $C_g |{\bf n }, {\bf m }\rangle$ 
display the same intra- and inter-species interactions, as shown in the example of Figure~\ref{fock2} 
and \ref{fock4}. Hence, we say that $|{\bf n }, {\bf m }\rangle$ and $C_g |{\bf n }, {\bf m }\rangle$ 
belong to the same ``configuration''. In what follows, when we mention  ``configuration" of Fock 
states we refer to the following formal definition. 

A {\em{configuration}} of Fock states in $\mathscr{H}^{a}\bigotimes\mathscr{H}^{b}$ is defined 
as a subset in the Fock basis such that all elements in the subset are obtained by acting on 
some fixed $|{\bf n }, {\bf m }\rangle$ with $C_g$, for {\em{all}} $g$ in $G$. In other 
words, a configuration groups together all Fock states which only differ in the location of 
individual bosons but with the {\em{same relative}} positions of bosons. According to this 
definition, the Fock basis can be partitioned in terms of distinguishable
configurations. These are equivalence classes~\cite{settheory} $[|{\bf n }, {\bf m }\rangle]$ 
defined by the property that any Fock states $|{\bf n }', {\bf m }'\rangle$ of the class
is related to the representative element $|{\bf n }, {\bf m }\rangle$
by the equivalence relation $|{\bf n }', {\bf m }'\rangle = C_g |{\bf n }, {\bf m }\rangle$ for some 
$g\in G$.
%
Then, the Fock basis is the disjoint union of configurations
%
$\bigcup_{k} [|{\bf n }_k, {\bf m }_k \rangle] $,
where the subscript $k$ runs through configurations.
It is useful to introduce in each class 
$[|{\bf n }_k, {\bf m }_k \rangle ]= 
\{ |{\bf n }^\alpha_k, {\bf m }^\alpha_k \rangle, 1\le \alpha \le n_k]\}$
the superscript $\alpha$ which enumerates the number $ n_k$ of Fock states belonging to the same configuration $k$. 
In terms of this partition, the Hilbert space $\mathscr H^{a}\bigotimes\mathscr{H}^{b}$ can be 
decomposed as a direct sum $\mathscr H^{a}\bigotimes\mathscr{H}^{b}=\bigoplus_k\mathscr{H}^{ab}_k$, 
where $\mathscr{H}^{ab}_k$ is the subspace spanned by the $k$th configuration.

Consider two arbitrary Fock states $|{\bf n }_k, {\bf m }_k \rangle$ and $|{\bf n }_k', {\bf m }_k' \rangle$ of the same configuration $k$, for which, of course, 
$|{\bf n }_k', {\bf m }_k' \rangle = C_g |{\bf n }_k, {\bf m }_k \rangle $ for some $g$. 
Based on the definition of $C_g$ and the symmetry of $|\Psi\rangle$ under the action of $\bf g$, 
we have $\langle {\bf n }_k', {\bf m }_k' |\Psi \rangle =
\langle {\bf n }_k, {\bf m }_k | C^+_g|\Psi \rangle
=
\langle {\bf n }_k, {\bf m }_k |\Psi \rangle$, 
being $C^+_g = C^{-1}_{g}= C_{g^{-1}}$. 
It follows that the ground state $|\Psi\rangle $ has the same expansion coefficients for all 
Fock states belonging to the same configuration. As a consequence, by projecting $|\Psi \rangle$
in the Fock basis, 
the ground state is only expanded on configurations rather than states, namely
\begin{equation}
\label{Eq1}
|\Psi \rangle
=\sum_k c_k |\chi_k\rangle \; , \qquad
|\chi_k \rangle \equiv \frac{1}{\sqrt{ n_k}} 
\sum_{\alpha=1}^{ n_k} |{\bf n }^\alpha_k, {\bf m }^\alpha_k \rangle\; ,
\end{equation}
where the class-dependent $|\chi_k \rangle$ have been defined.
Here, each $c_k$ is positive, and each $|\chi_k \rangle$ is normalized and spans an invariant 
subspace under the action of $G$, i.e. $C_g |\chi_k \rangle = |\chi_k \rangle$.
%
%
%
We denote this invariant subspace by $\mathfrak{h}^{ab}_k$. Since $|\chi_k \rangle$ is expanded 
by Fock states in the $k$th configuration, we have $\mathfrak{h}^{ab}_k\subset\mathscr{H}^{ab}_k$. 
Most importantly, Equation~\ref{Eq1} implies that $|\Psi \rangle$ lives in 
$\bigoplus_k\mathfrak{h}^{ab}_k$.
It is also important to notice that for any state 
$|\psi \rangle \in\mathscr H^{a}\bigotimes\mathscr{H}^{b}$, 
$|\psi \rangle$ is contained in $\bigoplus_k\mathfrak{h}^{ab}_k$ if and only if 
it is invariant under the action of $G$ or has the $G$-symmetry ($G$-symmetry of 
$|\Psi \rangle$ requires it being expanded equally on Fock states in each configuration). 
Moreover, since the Hamiltonian $H$ commutes with all $C_g$, then, for any 
$|\psi \rangle \in \bigoplus_k\mathfrak{h}^{ab}_k$ 
and $g\in G$, $C_g H |\psi \rangle= H C_g |\psi \rangle=H |\psi \rangle$, 
which implies that the direct sum $\bigoplus_k \mathfrak{h}^{ab}_k$ is also 
invariant under the action of $H$.
Therefore, for the purpose of studying the ground state and its energy, one can reduce 
the two-component Bose-Hubbard model to be defined in $\bigoplus_k\mathfrak{h}^{ab}_k$. 

\begin{figure}
\centering
\subfigure[\label{auto1}]{\includegraphics[width=0.10\textwidth]{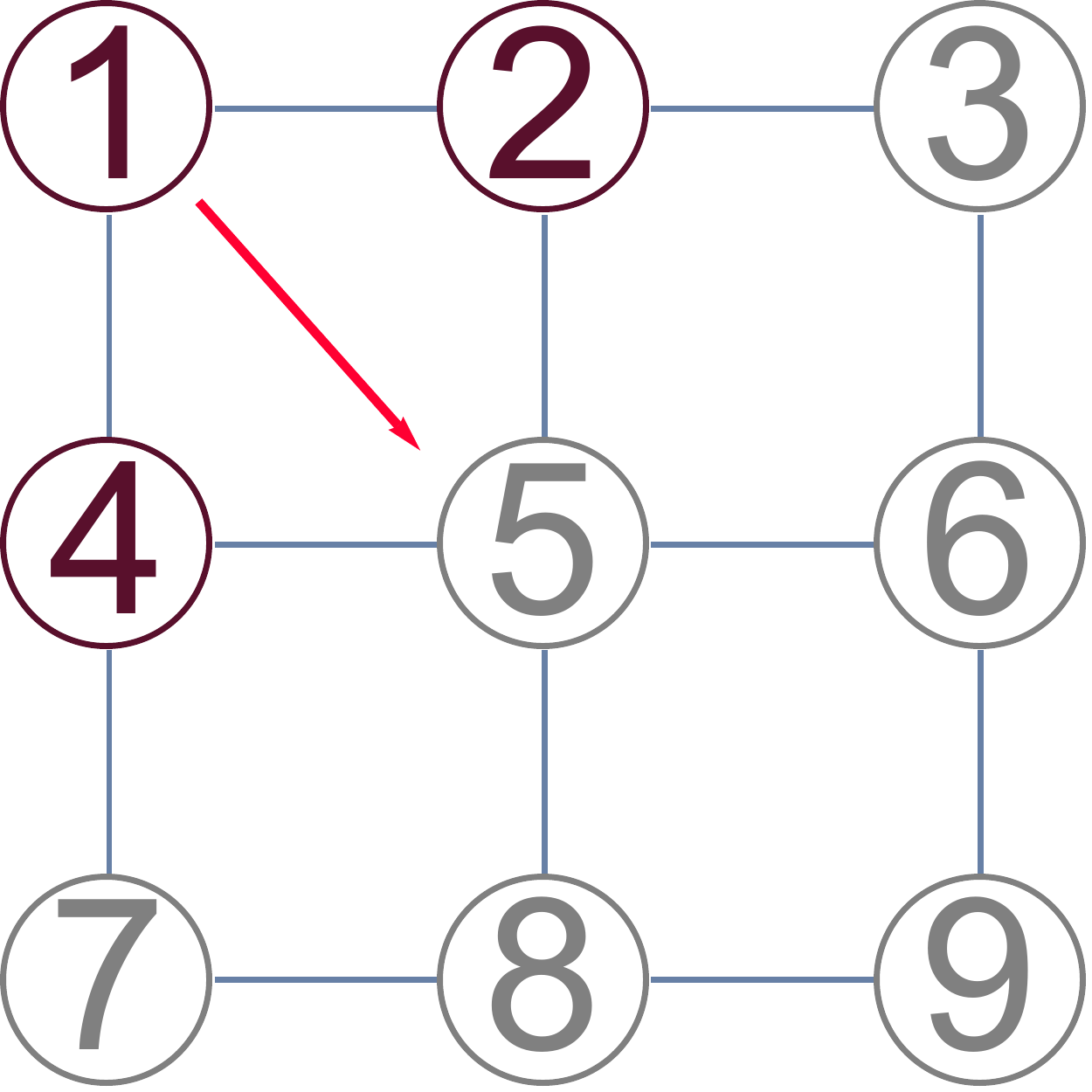}}

\subfigure[\label{auto2}]{\includegraphics[width=0.10\textwidth]{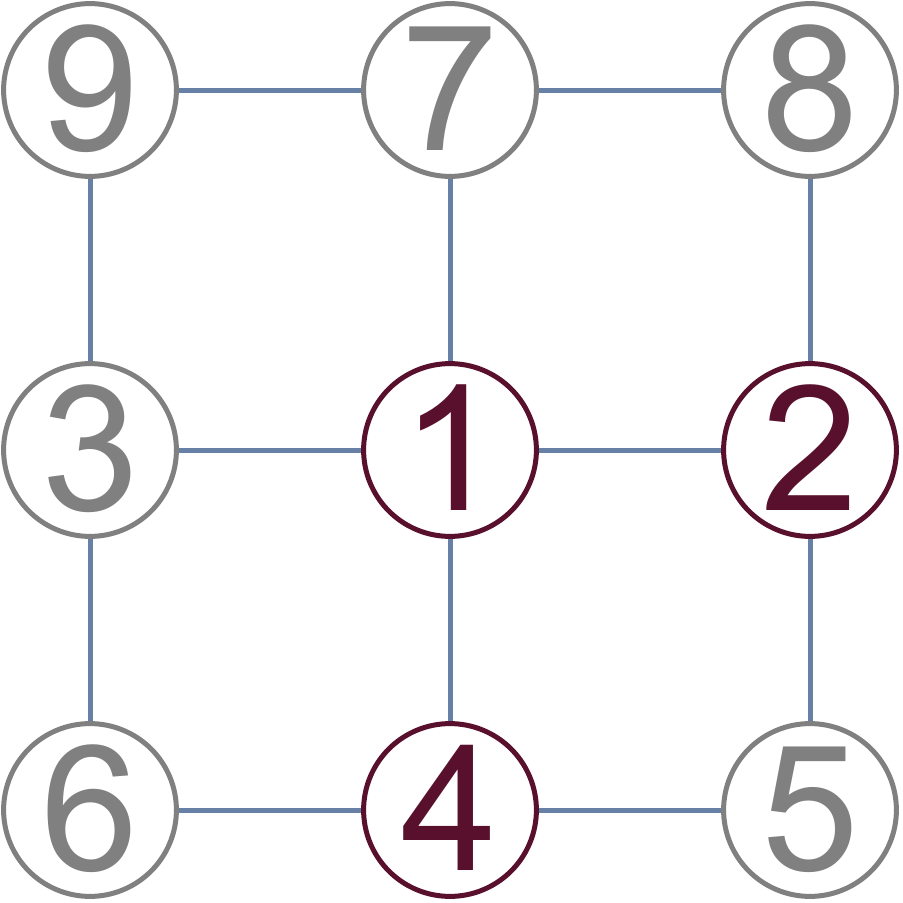}}\quad
\subfigure[\label{auto3}]{\includegraphics[width=0.10\textwidth]{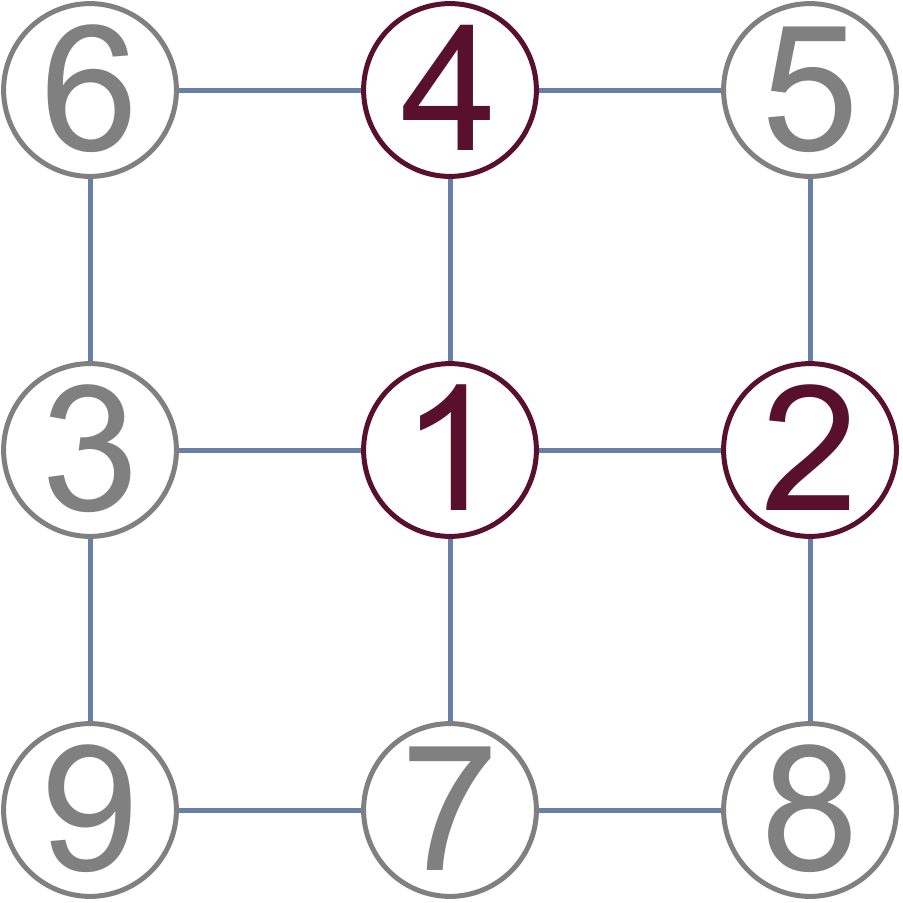}}\quad
\subfigure[\label{auto4}]{\includegraphics[width=0.10\textwidth]{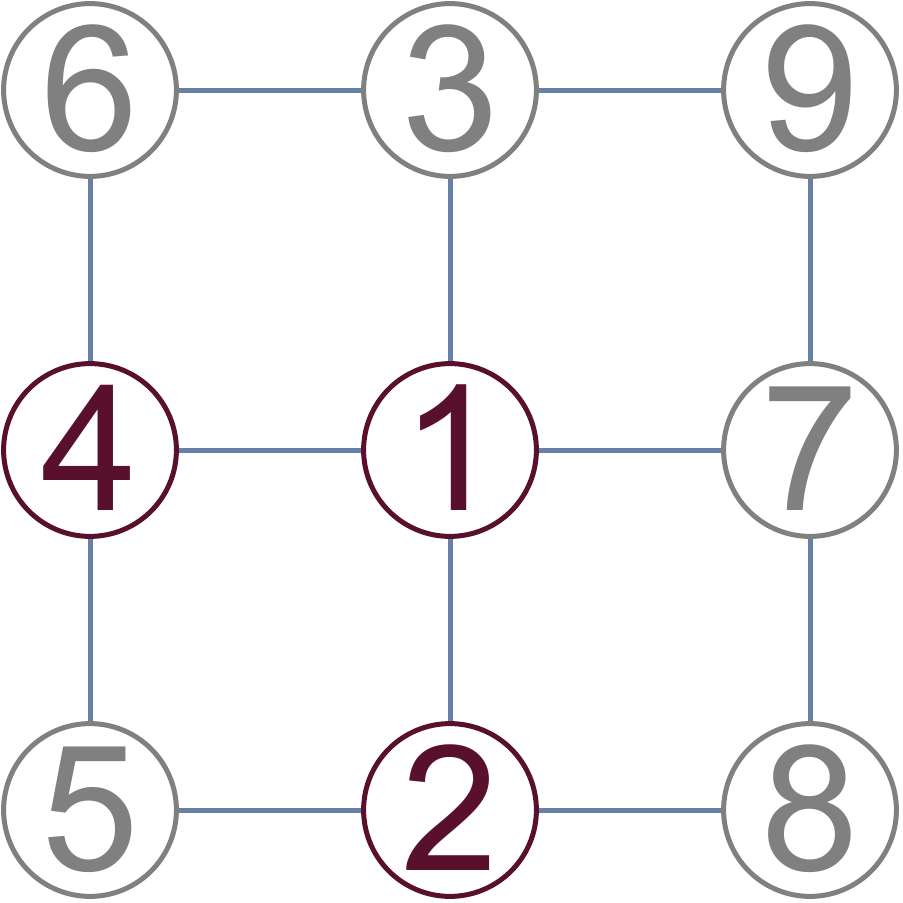}}\quad
\subfigure[\label{auto5}]{\includegraphics[width=0.10\textwidth]{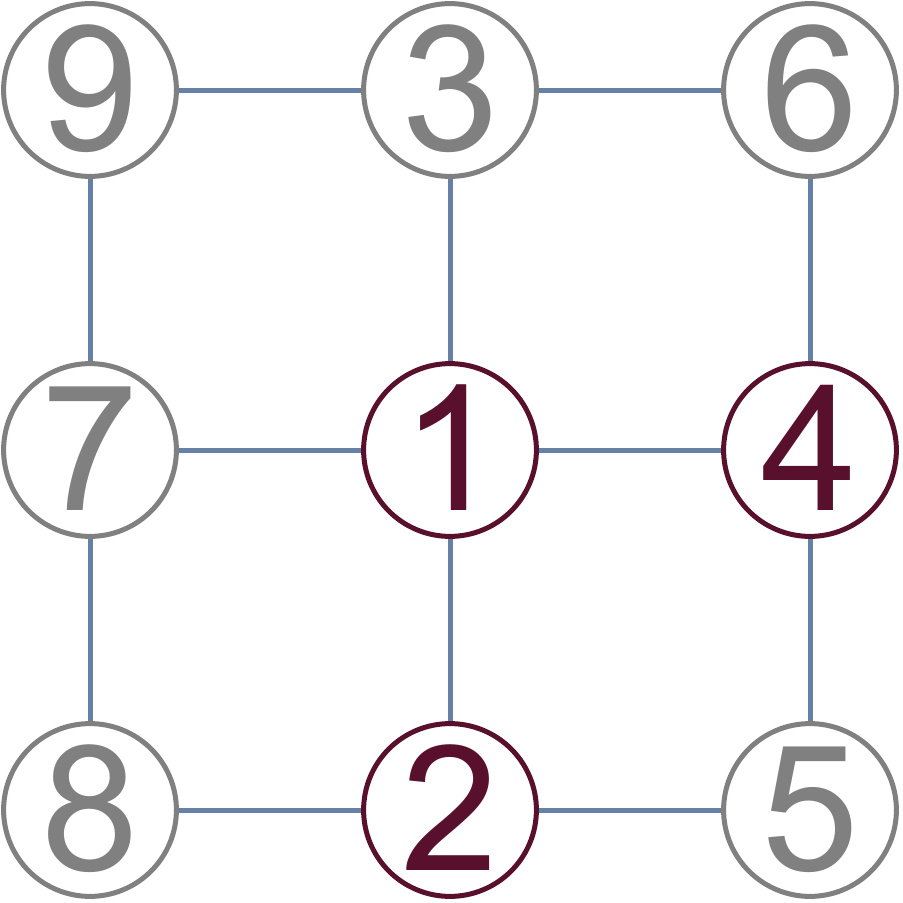}}\quad
\subfigure[\label{auto6}]{\includegraphics[width=0.10\textwidth]{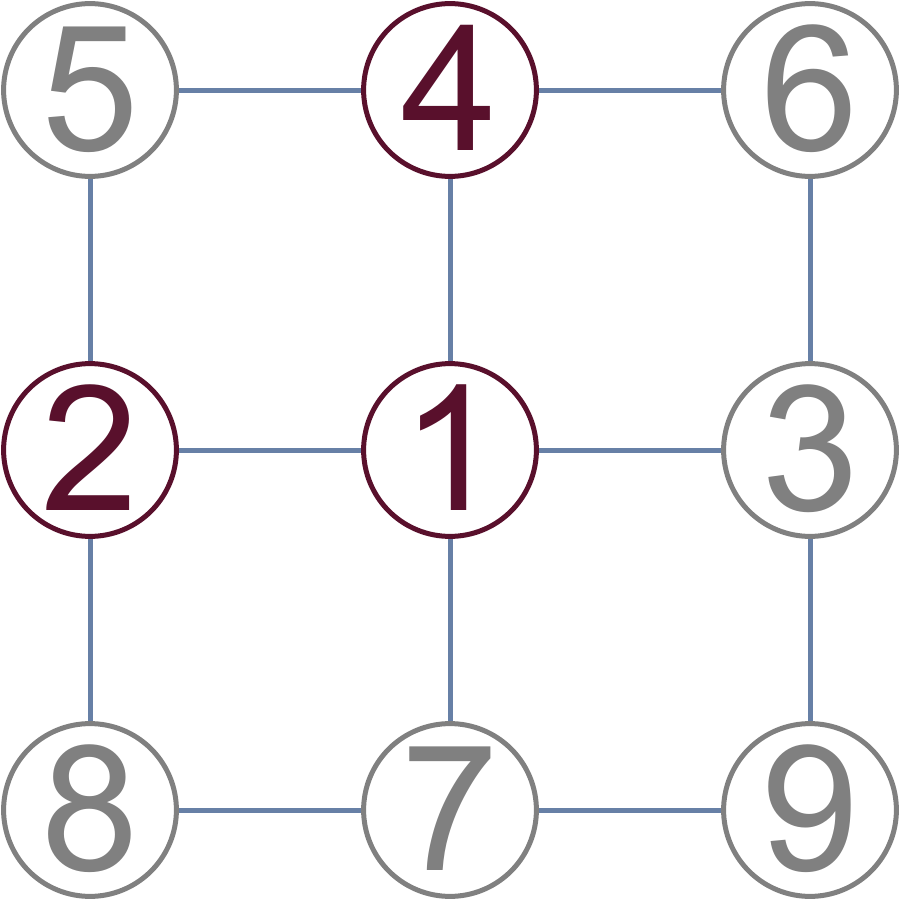}}\quad
\subfigure[\label{auto7}]{\includegraphics[width=0.10\textwidth]{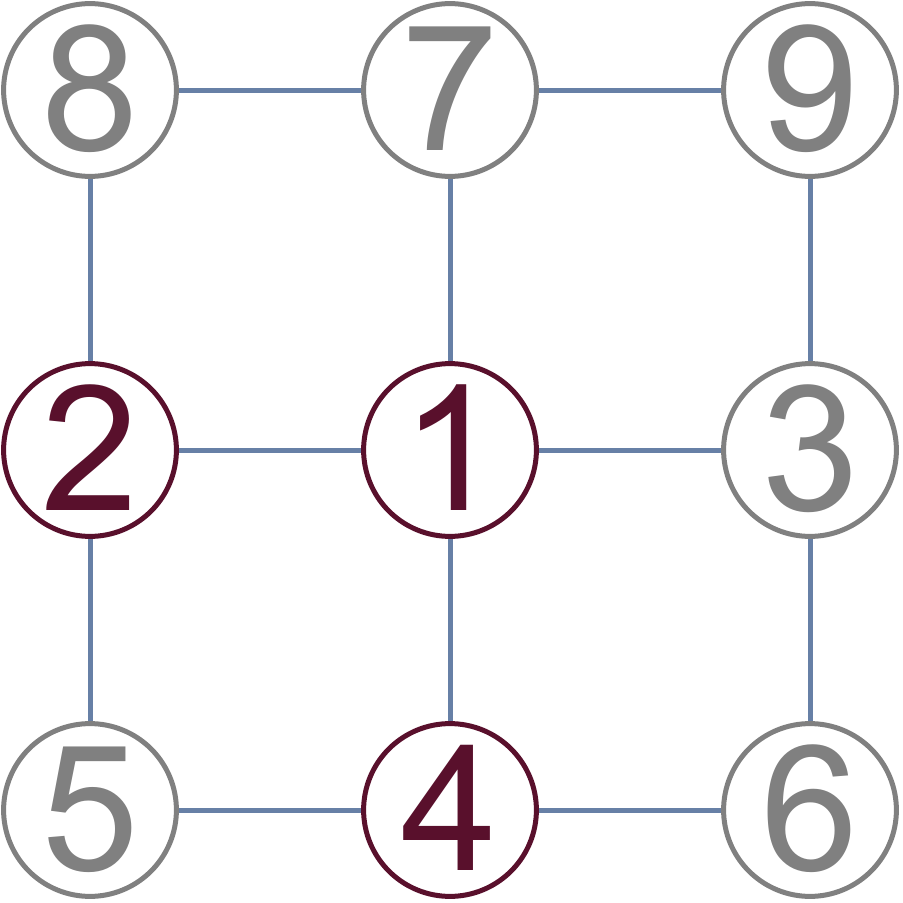}}\quad
\subfigure[\label{auto8}]{\includegraphics[width=0.10\textwidth]{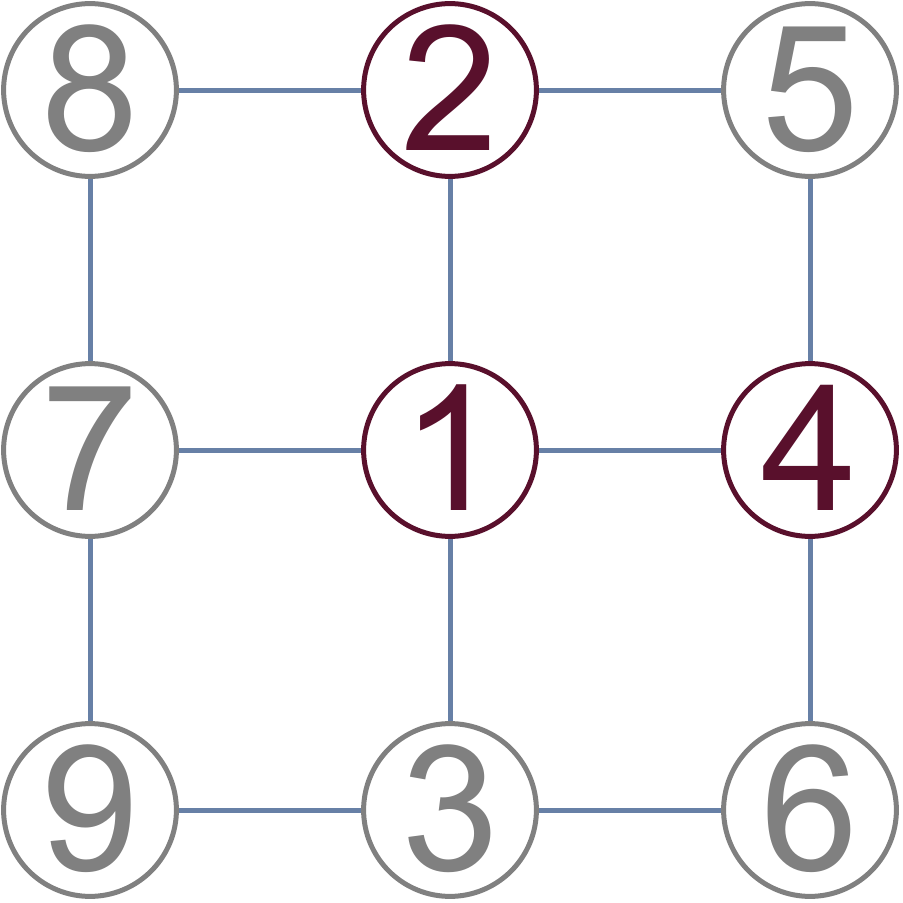}}\quad
\subfigure[\label{auto9}]{\includegraphics[width=0.10\textwidth]{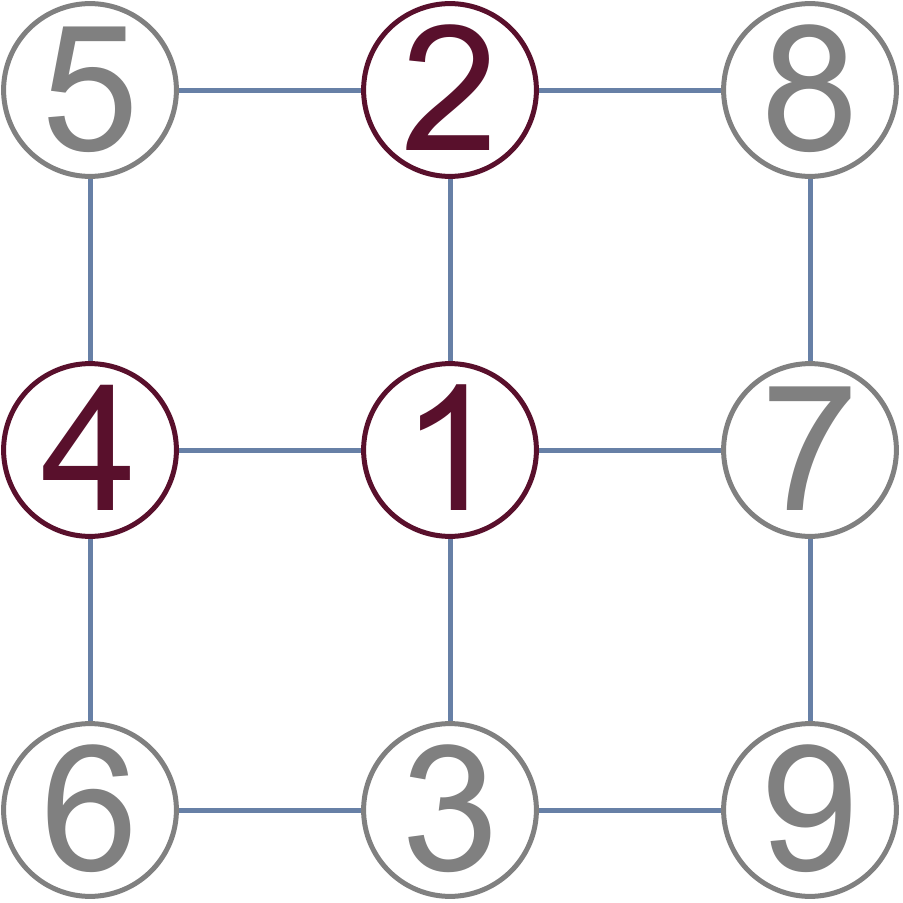}}
\caption{\label{FigThree}a) Sites 1,2 and 4 specify a coordinate system centered at position of site 1. 
b)-i) Eight graph automorphisms which specify the new positions of sites 2 and 4 while leaving the lattice unchanged (see text). }
\end{figure}

This is one of the central results of this paper. It demonstrates the property 
that quantum phase transitions of the two-component Bose-Hubbard model are only associated to 
the degrees of freedom describing the relative locations of bosons. Moreover, this reduction of 
the Hilbert space allows to greatly reduce the numerical cost of the perturbative calculation.
In particular, the matrix size of the Hamiltonian can be considerably reduced. 
For example, for a $ L\times  L$ square lattice with periodic boundary condition, the number of 
Fock states $n_k$ in each configuration can be as large as $8\times  L^2$. Referring to 
Figure~\ref{auto1}, all graph automorphisms can be obtained by doing the following. 
Consider sites, e.g., 1, 2 and 4, as specifying a coordinate system centered at position of site 1. 
This coordinate system can be arbitrarily mapped onto another coordinate system originating 
at any other site of the lattice and such that the three sites 1, 2 and 4 maintain their 
relative position. This gives a factor of $L^2$. Next, for a fixed origin of the coordinate 
system (position of site 1), there exist 8 choices for mapping sites 2 and 4 using rotations 
and reflections which leave the lattice unchanged (see Figures~\ref{auto2}-\ref{auto9}). 
Therefore, in this example, there are $8\times  L^2$ graph automorphisms in total.
As a consequence, the matrix size can be roughly reduced by a factor $1/(8\times  L^2)^2$.

%
\section{The criterion of inter-species entanglement}
\label{sec5}
In this Section we investigate how the inter-species entanglement arises in the ground state of a binary interacting mixture. As discussed above, we will work on a reduced Hilbert space.

The inter-species interaction characterized by a strength $U^{ab}$ is the source of the 
inter-species entanglement. This is because in the noninteracting binary mixture the Hamiltonian 
is simply the sum of two single-species Bose-Hubbard models, and therefore, due to the uniqueness 
of the ground state one has $|\Psi \rangle =|\Phi^a \rangle \otimes |\Phi^b \rangle $ and thus 
$|\Psi \rangle $ is non-entangled. Here $|\Phi^a \rangle $ and $|\Phi^b \rangle $ are unique 
ground states of the single-component systems defined in $\mathscr H^a$ and $\mathscr H^b$, 
respectively (the uniqueness of these ground states can be shown by similar arguments used 
in Reference~\cite{wei1}). Therefore, the inter-species interaction is a {\em{necessary}} 
condition for finite entanglement of binary bosonic mixtures in optical lattices.

To further understand the role played by the inter-species interaction, we start by studying 
the ground state of noninteracting mixtures and its symmetry properties. 
By applying similar arguments as in Reference~\cite{wei1}, we can conclude that 
$\langle {\bf n} |{\Phi^a} \rangle  >0$ and $\langle {\bf m} |{\Phi^b} \rangle  >0$ 
for the Fock bases $\{ | {\bf n} \rangle \}$ and $\{|{\bf m} \rangle\}$ defined in 
$\mathscr{H}^a$  and  $\mathscr{H}^b$ respectively. Then, the above analysis on 
representation map $C$ can be applied to both $A$ and $B$. 
For this reason, we can define subspaces $\mathfrak{h}^{a}_i\subset\mathscr{H}^a_i$
and  $\mathfrak{h}^{b}_j\subset\mathscr{H}^b_j$ corresponding to the $i$th and the 
$j$th  configurations of the single-species Fock bases $\{| {\bf n} \rangle \}$ and 
$\{|{\bf m} \rangle \}$, in a similar way as we defined 
$\mathfrak{h}^{ab}_k\subset \mathscr{H}^{ab}_k$. Likewise, we have 
$$
|\Phi^a \rangle \in {\bigoplus}_i\mathfrak{h}^{a}_i\subset {\bigoplus}_i\mathscr{H}^a_i
= \mathscr{H}^a\, , \quad {\rm and }\,\,\,
| \Phi^b\rangle \in {\bigoplus}_j\mathfrak{h}^{b}_j\subset {\bigoplus}_j\mathscr{H}^b_j
=\mathscr{H}^b \; .
$$ 
Based on the symmetries of $|\Phi^{a} \rangle$ and $|\Phi^{b}\rangle$, it follows that 
$A_g\otimes B_{g'} |\Psi \rangle
= A_g | \Phi^a \rangle \otimes B_{g'} |\Phi^b \rangle
= |\Phi^a \rangle \otimes |\Phi^b \rangle = |\Psi\rangle$ for any two graph automorphisms 
$g$ and $g'$. Considering that $A \otimes B : (g,g') \mapsto A_g \otimes B_{g'}$ defines a 
unitary representation of the direct product group $G \ast G$ in which the 
number of group elements is squared, we conclude that $| \Psi\rangle $ is invariant under 
the action of the direct product of the graph automorphism group. In other words, in the 
absence of inter-species interaction, the system is invariant when two arbitrary graph 
automorphisms $g$ and $g'$ operate on the two species independently. On the other hand, 
for $U^{ab}\ne0$, the system is generally {\em{not}} invariant under the action of $g$ and $g'$ 
independently, unless $g=g'$. This implies that the absence of the inter-species interaction 
`loosens' the restrictions on the operations which leave the ground state invariant making 
the system more symmetric. Indeed, the lack of $G \ast G$ symmetry 
will be shown to serve as a criterion for the appearance of entanglement in the 
ground state $|\Psi \rangle$ once the inter-species interaction is turned on.

We now turn on the inter-species interaction and look at the decomposition 
of the Hilbert spaces in order to gain further insight in the structure of $|\Psi \rangle$. 
First, we consider the $i$th configuration of species-a and the $j$th configuration of species-b, 
i.e. $\{ |{\bf n}^1_i \rangle, |{\bf n}^2_i \rangle, ..., |{\bf n}^{r_i}_i \rangle \}$ 
and  $\{ |{\bf m}^1_j \rangle,|{\bf m}^2_j \rangle, ..., |{\bf m}^{s_j}_j \rangle \}$
with labels interpreted as for the configuration of the mixture, that is, subscripts refer 
to configurations while superscripts refer to Fock states within a configuration. We define 
the product-configuration $(i,j)$ as 
$$
\{
|{\bf n}^{1}_i \rangle |{\bf m}^{1}_j \rangle, 
|{\bf n}^{1}_i \rangle |{\bf m}^{2}_j \rangle, ... , 
|{\bf n}^{2}_i \rangle|{\bf m}^{1}_j \rangle ,
|{\bf n}^{2}_i \rangle |{\bf m}^{2}_j \rangle, ...,
|{\bf n}^{r_i}_i \rangle|{\bf m}^{s_j}_j \rangle
\}, 
$$
i.e. the collection of all possible products of Fock states from the $i$th configuration of species-a and the $j$th configuration of species-b. In terms of the definition of configurations for single species, one can alternatively define the product-configuration by fixing some Fock state 
$|{\bf n}, {\bf m} \rangle = |{\bf n} \rangle \otimes |{\bf m} \rangle$ 
and collecting all $A_{g_1} |{\bf n} \rangle \otimes B_{g_2} |{\bf m} \rangle$ 
running through all $g_1,g_2\in G$.

\begin{figure}
\centering
\subfigure[\label{confa}]{\includegraphics[width=0.17\textwidth]{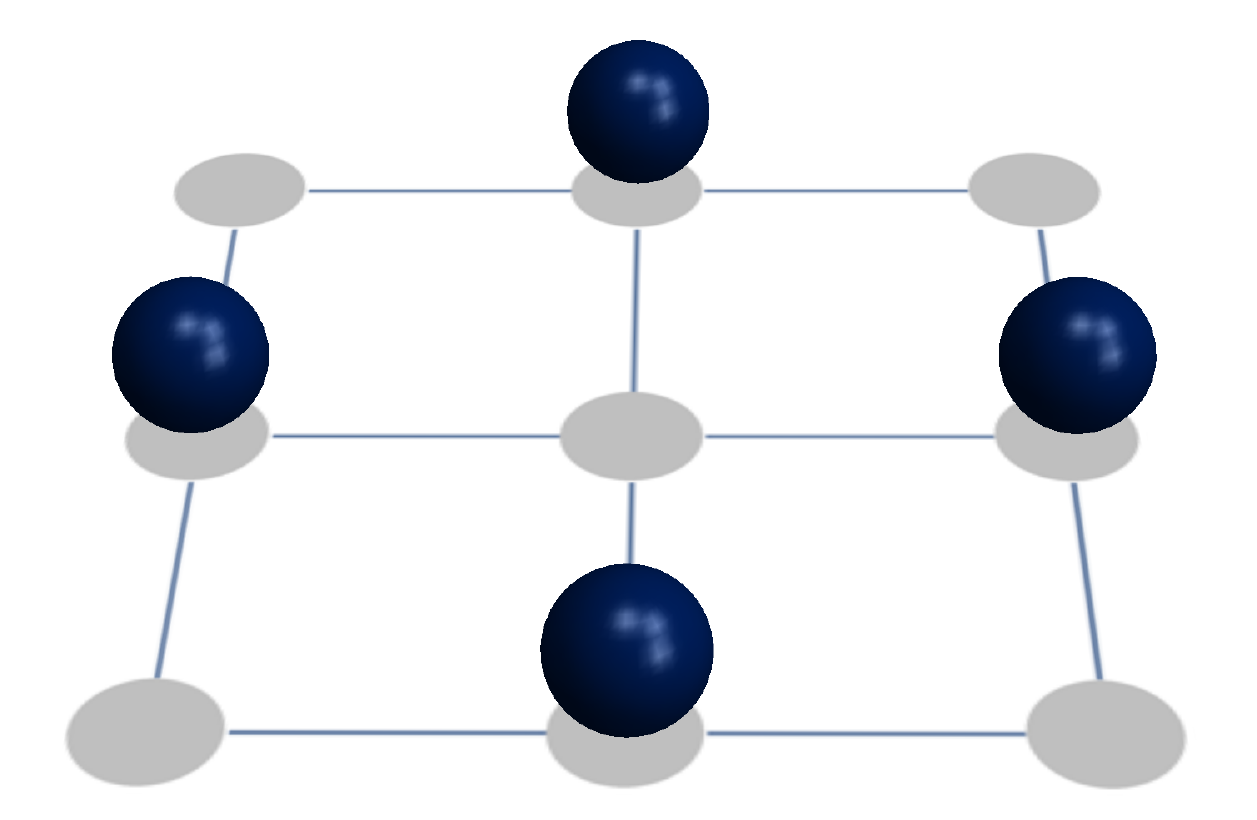}}\quad
\subfigure[\label{confb}]{\includegraphics[width=0.17\textwidth]{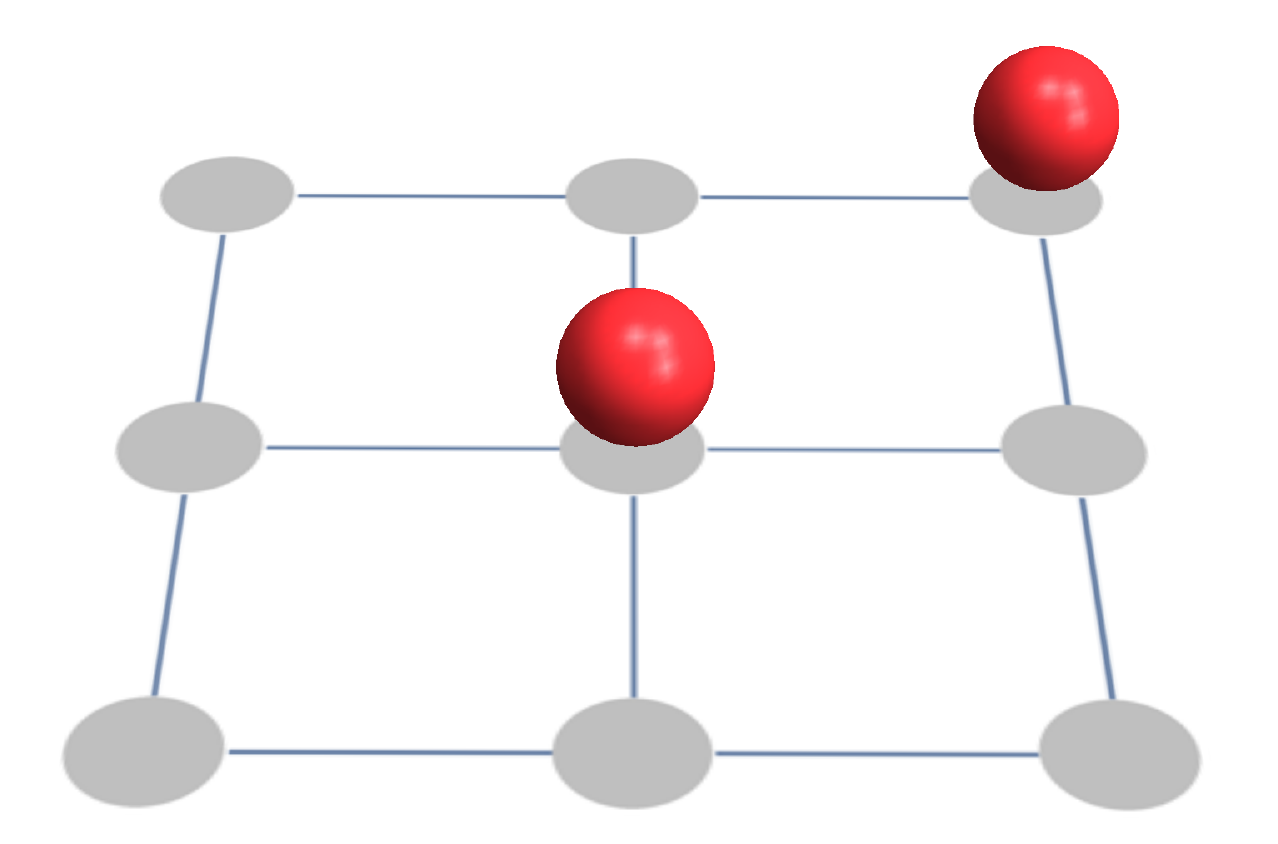}}\quad
\subfigure[\label{conf1}]{\includegraphics[width=0.17\textwidth]{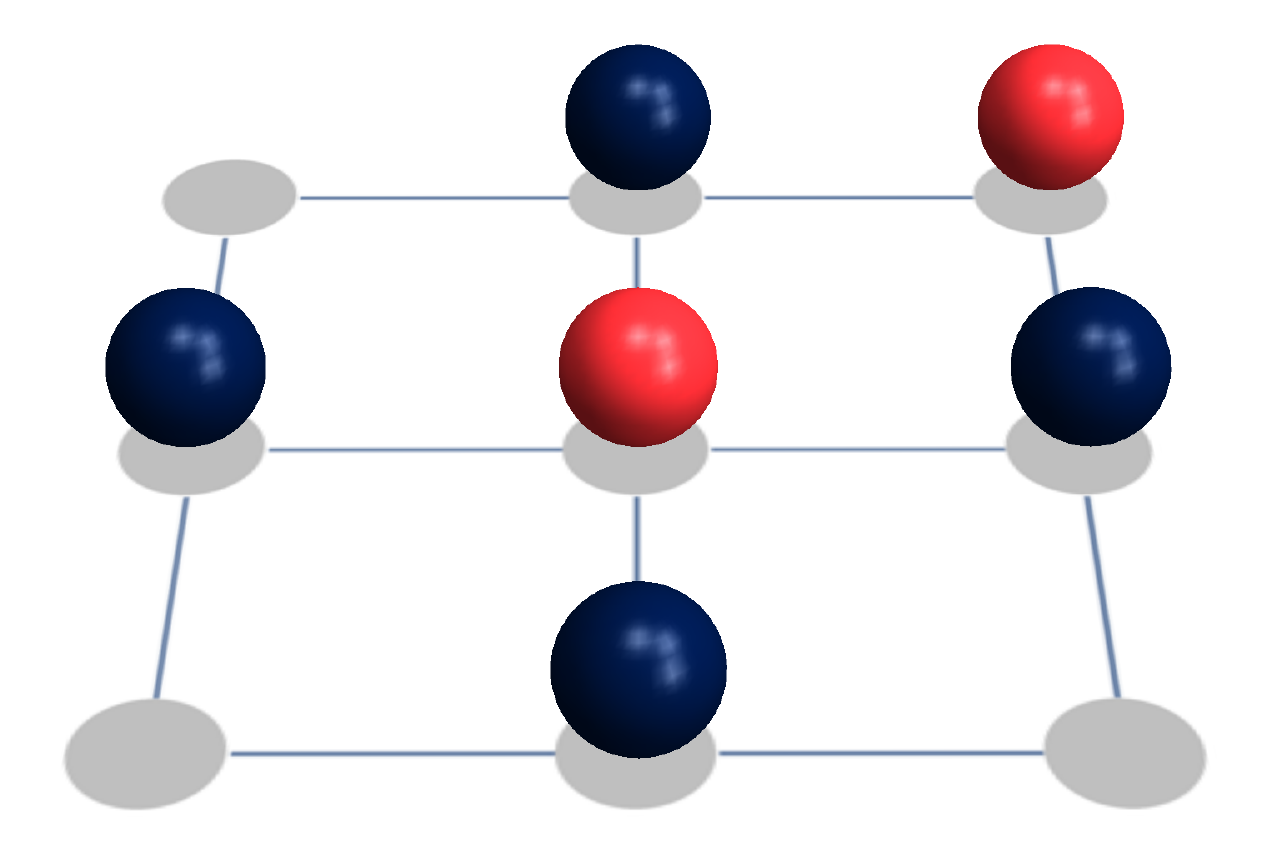}}\quad
\subfigure[\label{conf2}]{\includegraphics[width=0.17\textwidth]{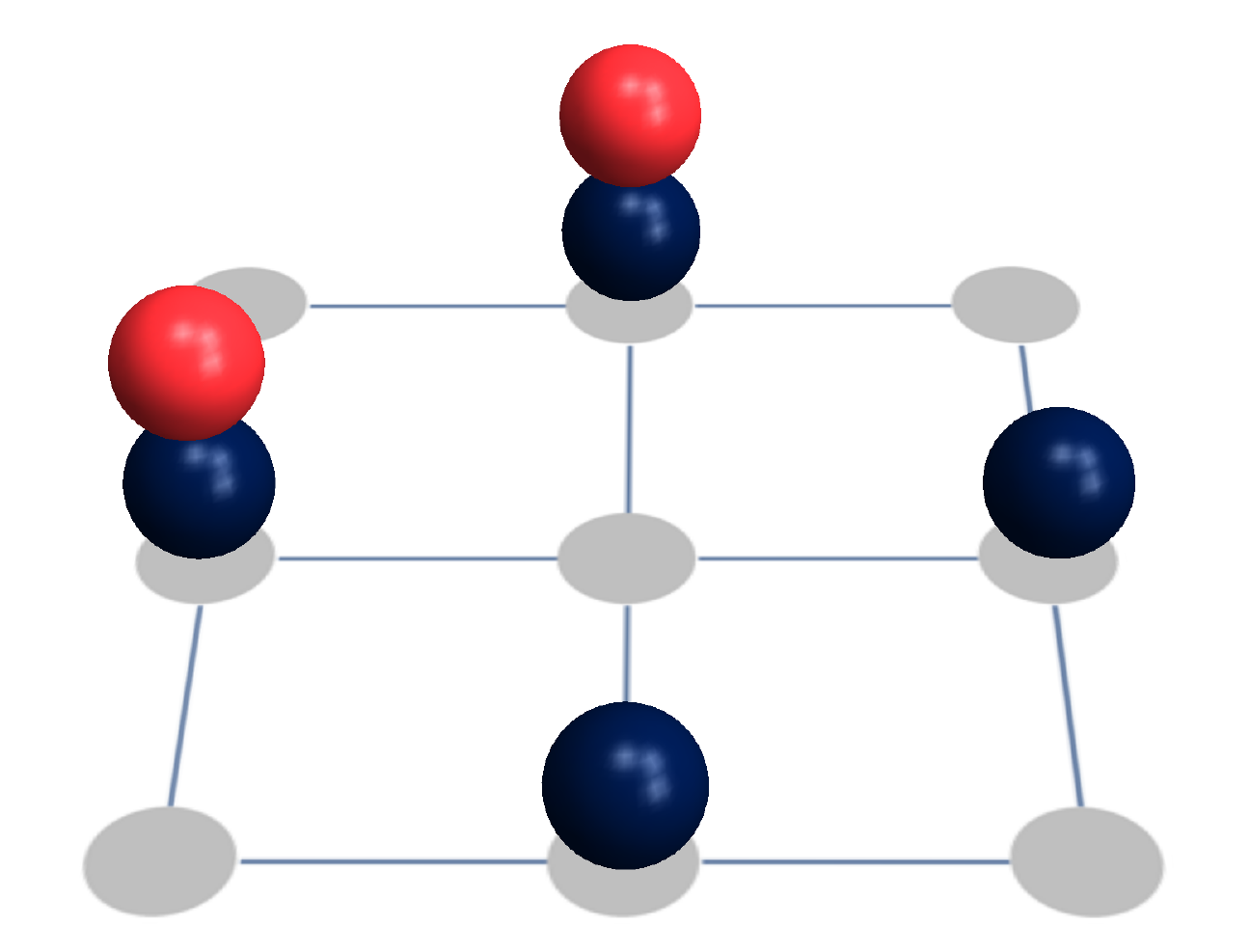}}
\caption{\label{FigFour}a) and b) Example of a configuration of species-a and of species-b, respectively. c) and d) Configurations of the mixture contained in the product-configuration resulting from single species configurations sketched in a) and b).}
\end{figure}

Let us consider an arbitrary product-configuration $(i,j)$ and fix a Fock state 
$|{\bf n}, {\bf m} \rangle $ in this configuration with $|{\bf n} \rangle$ belonging 
to the $i$th configuration of species-a and $|{\bf m} \rangle $ belonging to the $j$th 
configuration of species-b. If we choose two arbitrary $g_1$ and $g_2$ from $G$, and 
collect all 
$ A_g A_{ g_1} |{\bf n}\rangle 
\otimes
B_{g} B_{g_2 }| {\bf m} \rangle 
= A_{g g_1} |{\bf n} \rangle 
\otimes B_{g g_2 } |{\bf m} \rangle 
$ 
with $g$ running through all $g\in G$, we obtain a configuration of the mixture which is surely 
included in the product-configuration $(i,j)$. Then, we repeat this process by choosing two other 
$g_3$ and $g_4$ from $G$ with $ A_{ g_3} |{\bf n}\rangle \otimes B_{g_4 }| {\bf m} \rangle $
not included in the previously obtained configuration, and we obtain another configuration 
of the mixture which is disjoint from  the former one. Thus, inductively, we can partition 
the product-configuration $(i,j)$ in terms of configurations of the mixture (below we refer to 
configurations of the mixture simply as `configurations', as defined in Section~\ref{sec4}, 
as opposed to product-configurations defined above). In other words, the product-configuration 
$(i,j)$ is the union of the configurations included in it. 
This is displayed in Figure~\ref{FigFour}, where the configurations of species-a and species-b, 
shown in \ref{confa} and \ref{confb} respectively, determine a product-configuration which consists 
of the union of the two configurations displayed in \ref{conf1} and \ref{conf2} (note that 
we only display a single Fock state per configuration).

Obviously, each configuration has to be included in some product-configuration because both configurations and product-configurations partition the same Fock basis. Let us denote the $k$th configuration included in the product-configuration $(i,j)$ by $k\in(i,j)$. Then, according to the definition of $\mathscr{H}^a_i$, $\mathscr{H}^b_j$ and $\mathscr{H}^{ab}_k$, we have 
$$
\mathscr{H}^a_i\bigotimes\mathscr{H}^b_j=\bigoplus_{k\in(i,j)}\mathscr{H}^{ab}_k
\; .
$$
%
%
%
Furthermore, since $\mathfrak{h}^a_i$ and $\mathfrak{h}^b_j$ are spanned by 
$| \zeta_i \rangle =(1/\sqrt{ r_i})\sum_{\alpha=1}^{ r_i} |{\bf n}_i^\alpha \rangle$ 
and $| \lambda_j \rangle  =(1/\sqrt{ s_j})\sum_{\beta=1}^{ s_j} |{\bf m }_j^\beta \rangle$, 
respectively, then $\mathfrak{h}^a_i\bigotimes\mathfrak{h}^b_j$ is spanned by 
$$
| \zeta_i \rangle \otimes | \lambda_j \rangle
= \frac{1}{\sqrt{ r_i s_j} } \sum_{\alpha = 1}^{ r_i}\sum_{\beta=1}^{s_j}
|{\bf n}_i^\alpha \rangle \otimes |{\bf m}_j^\beta \rangle
$$
i.e. the equally weighted sum of the Fock states contained in the product-configuration 
$(i,j)$. Moreover, for any $k\in(i,j)$, 
$| \chi_k \rangle =(1/\sqrt{ n_k})\sum_{\sigma=1}^{ n_k}
|{\bf n}_k^\sigma \rangle \otimes |{\bf m}_k^\sigma \rangle
$
is also an equally weighted sum of a subset of states in the product-configuration $(i,j)$ 
which defines the $k$th configuration. Here, every 
$|{\bf n}_k^\sigma \rangle \otimes |{\bf m}_k^\sigma \rangle$ is equal to some 
$|{\bf n}_i^\alpha \rangle \otimes |{\bf m}_j^\beta \rangle$ in the product-configuration 
(to avoid confusion, we use subscript $k$ to indicate configurations of the mixture and 
subscripts $i$ and $j$ to indicate configurations of species-a and species-b respectively). 
Since product-configuration $(i,j)$ consists of all $k$th configurations with $k\in(i,j)$, 
then $|\zeta_i \rangle \otimes |\lambda_j\rangle 
=(1/\sqrt{ r_i s_j})\sum_{k \in(i,j)}\sqrt{ n_k} |\chi_k \rangle$. 
The latter equality implies  
$\mathfrak{h}^a_i\bigotimes\mathfrak{h}^b_j\subset\bigoplus_{k\in(i,j)}\mathfrak{h}^{ab}_k$, 
and thus 
$$
\bigoplus_{i,j}(\mathfrak{h}^a_i \bigotimes \mathfrak{h}^b_j) \subset\bigoplus_{k}\mathfrak{h}^{ab}_k
\; .
$$
%
%
It is easy to show that a state belongs to the subspace 
$$
{\bigoplus}_{i,j} \left (\mathfrak{h}^a_i \bigotimes \mathfrak{h}^b_j \right )
=\left (
{\bigoplus}_i\mathfrak{h}^{a}_i \right )\bigotimes \left ( {\bigoplus}_j\mathfrak{h}^{b}_j \right )
$$ 
{\em{if and only if}} it is invariant under the action of the direct product group 
$ G \ast G$ (see~\ref{app2}). This result
%
%
is an essential property since, by applying the finiteness of the automorphism group, we can show 
that if the ground state $|\Psi \rangle$ is non-entangled, then 
$|\Psi \rangle \in \bigoplus_{i,j}(\mathfrak{h}^a_i \bigotimes \mathfrak{h}^b_j)$ 
as proved in \ref{app1}. 
Consequently, a non-entangled $\Psi$ implies that it is invariant under the action of any direct product of graph automorphisms $(g,g')\in G \ast G$. Alternatively, breaking the $G\ast G$ symmetry of $\Psi$ implies onset of entanglement of the mixture in the ground state. 
It is worth noting that in the case where there is only one boson of the second component, i.e. $ N^b=1$, the loss of $G\ast G$ symmetry of $\Psi$ serves as a necessary and sufficient condition for the onset of entanglement.
In this case, in fact, the second species has only one configuration. If $| \Psi \rangle$ is symmetric under 
$G \ast G$, then $| \Psi \rangle \in(\bigoplus_i\mathfrak{h}^{a}_i)\otimes\mathfrak{h}^{b}$. Since 
$\mathfrak{h}^{b}$ is one-dimensional, $| \Psi \rangle$ will always has the form 
$| \phi^a \rangle\otimes |\phi^b\rangle$, where $|\phi^b\rangle$ spans $\mathfrak{h}^{b}$, and 
as such $| \Psi \rangle$ will be non-entangled.

It would be interesting 
to investigate if, in the general case of 
$N^b\ne 1$, $|\Psi\rangle\in\bigoplus_{i,j}(\mathfrak{h}^a_i\bigotimes\mathfrak{h}^b_j)$ 
also implies that $|\Psi\rangle$ is non-entangled. If this is the case, then, the fact that the 
ground state belongs to $\bigoplus_{i,j}(\mathfrak{h}^a_i\bigotimes\mathfrak{h}^b_j)$ or, 
equivalently, as we have just shown, that the ground state lacks $G \ast G$ symmetry, is
a sufficient and necessary condition for the entangling of $|\Psi\rangle$. On the other hand, if 
this is {\em{not}} the case, then, the entangled 
ground state $|\Psi\rangle$ belongs to $\bigoplus_{i,j}(\mathfrak{h}^a_i\bigotimes\mathfrak{h}^b_j)$ and 
possesses $G \ast G$ symmetry. A finite entanglement implies a non-zero inter-species 
interaction $U^{ab}$ (recall we have shown that non-zero $U^{ab}$ is a necessary condition 
for an entangled ground state $|\Psi\rangle$). As $|U^{ab}|$ is increased and provided all other 
model parameter are kept fixed, the expansion coefficients of $|\Psi\rangle$ inside each 
product-configuration will eventually no longer be the same since Fock states which 
minimize the inter-species interaction energy will be favored, i.e. will have a 
greater weight. Thus, $|\Psi\rangle$ will no longer belong to 
$\bigoplus_{i,j}(\mathfrak{h}^a_i\bigotimes\mathfrak{h}^b_j)$. The implication of the 
above argument is that there would be a phase transition between a phase characterized by 
$G \ast G$ symmetry and non-zero $U^{ab}$ and a phase with {\em{broken}} $G \ast G$ symmetry. 
We are not attempting to answer this question here as this goes beyond the scope of the 
present work but we find this possibility intriguing and therefore worth it to mention.

\section{Numerical results}
\label{sec6}

In this Section, we discuss our results for the entanglement entropy of the ground state and the excited states corresponding to the hole- and particle-side. The entanglement entropy is the standard measure of bipartite entanglement for pure states. It is defined as the von Neumann entropy $\mathbf{S}$ of the reduced density operator $\rho^a=\Tr^b[ \rho]$, or $\rho^b=\Tr^a[\rho]$ where 
$\rho=|\Psi\rangle\langle\Psi|$~\cite{entanglementbook2}.
In the following, we use the notation $\mathbf{e}(\rho)=\mathbf{S}(\rho^a)=\mathbf{S}(\rho^b)$.

\begin{figure}
\centering
\subfigure[\label{lobe1}]{\includegraphics[width=0.44\textwidth]{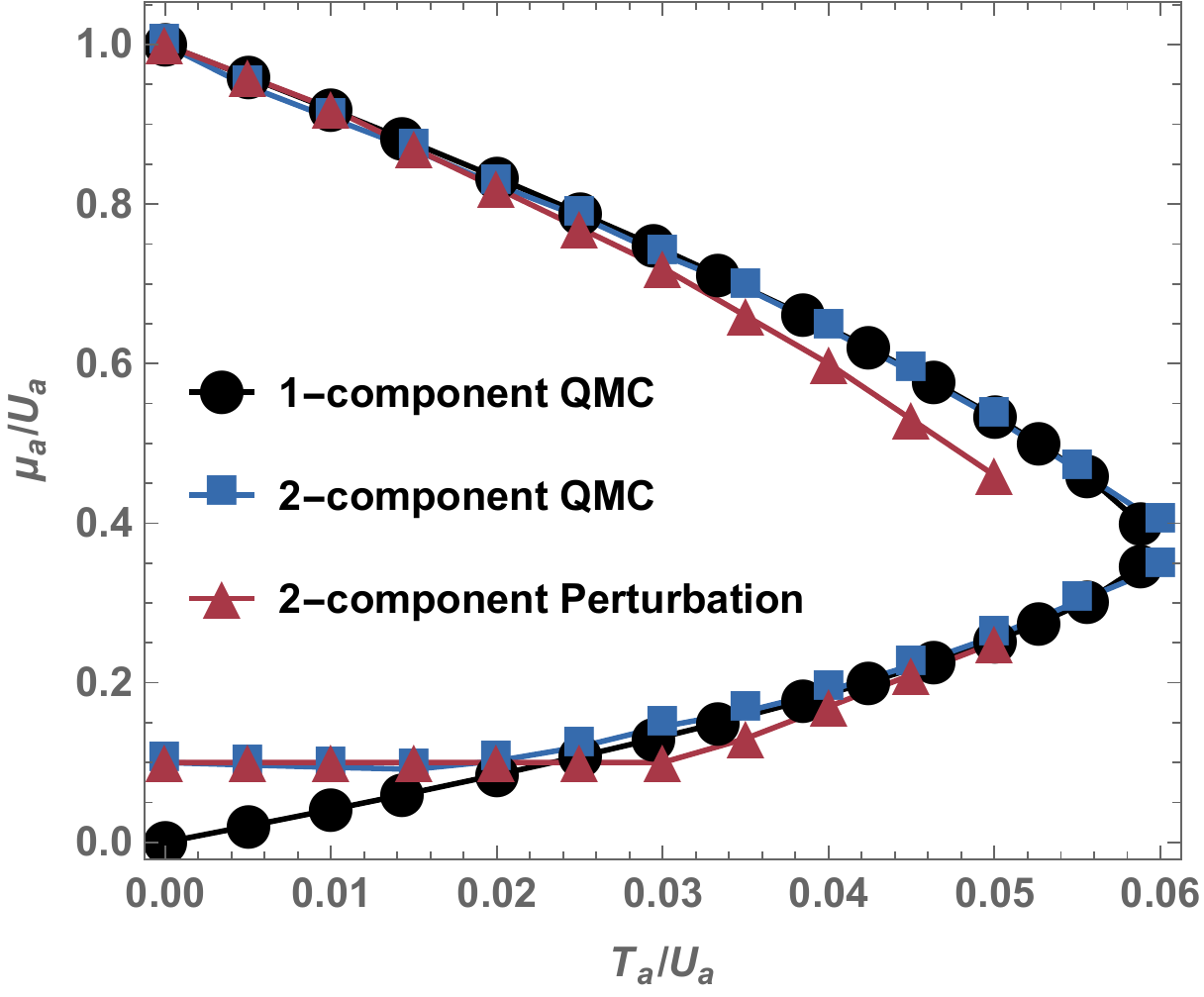}}\quad
\subfigure[\label{ee}]{\includegraphics[width=0.38\textwidth]{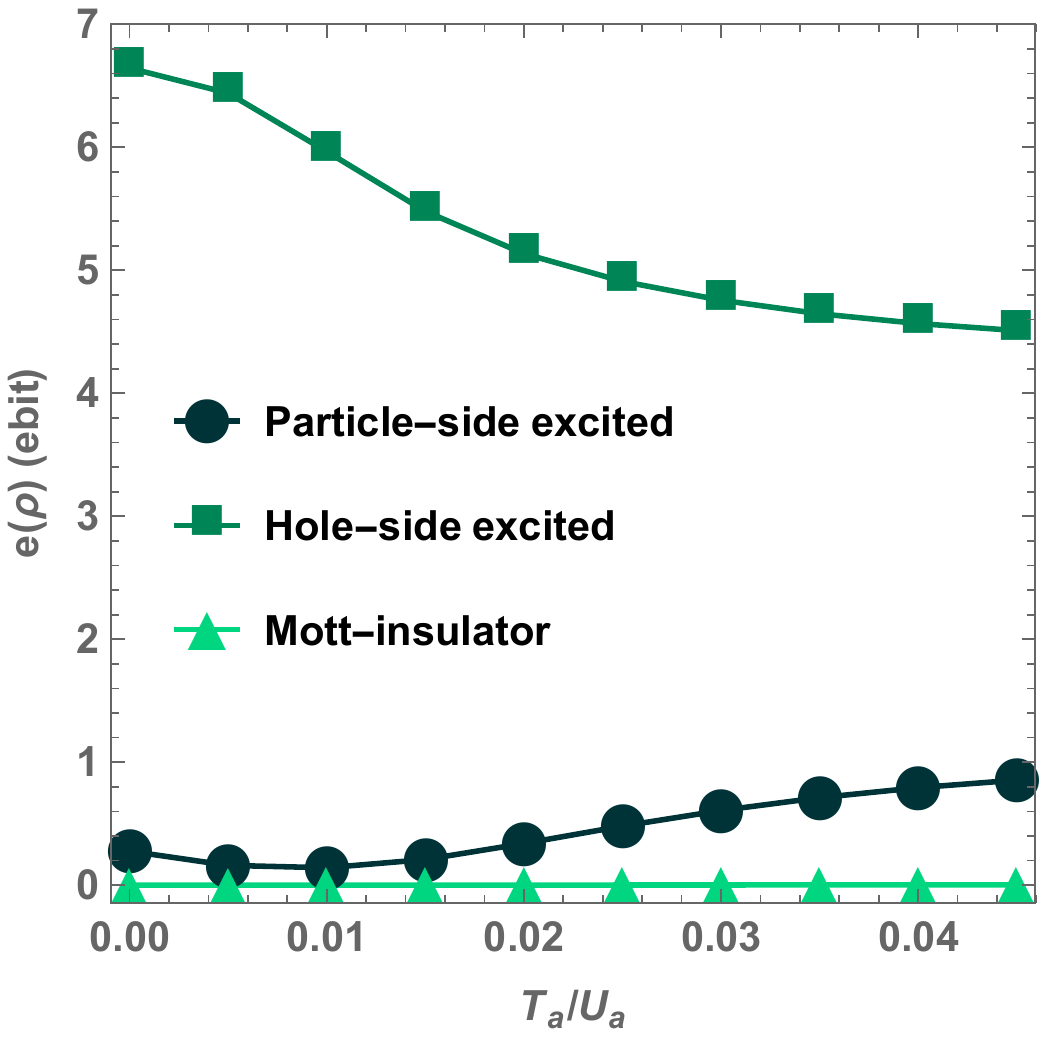}}
\caption{\label{FigFive}a) lobe boundaries of component-a in the presence of a single particle-b as computed with quantum Monte Carlo (blue squares) and by means of perturbation theory (red triangles). For comparison, we also plot the lobe boundaries of the single species Bose-Hubbard model (black circles) as computed with quantum Monte Carlo. b) entanglement entropies of the ground state (triangle), hole-side excited state (square) and particle-side excited state (circle) calculated using perturbation theory.}
\end{figure}

Our results are based on perturbation theory carried out on a $10\times 10$ square lattice with periodic boundary conditions. We have checked finite size effects for systems of linear size $L=6,7,8,9,10$ and found no sizable discrepancy on the lobe boundaries. The perturbative calculation treats the hopping term in Equation~\ref{Eq0} as the perturbation. The results presented refer to $T^a=T^b$, $U^{ab}=0.1U^a=0.1U^b$, and $N^b=1$. The validity of the perturbative calculation can be inferred by a comparison with quantum Monte Carlo (QMC) results by the two-worm algorithm~\cite{twoworms} of the first MI lobe of component-a. 
Figure~\ref{lobe1} shows the lobe boundaries as computed with QMC (blue squares) and by means of perturbation theory (red triangles). For comparison, we also plot the lobe boundaries of the single species Bose-Hubbard model (black circles) as computed with QMC. Overall, we see that the particle-side of the lobe (upper boundary) in the presence of a single particle-b is basically unaffected and lies on top of the boundary for the single-species case. On the other hand, at low hopping, the hole-side of the lobe (lower boundary) is prominently different from the single-species lobe boundary. This difference is more pronounced for smaller hopping, where quantum fluctuations are less important. As the ratio $T^a/U^a$ reaches a value of 0.03, we start seeing discrepancy between the perturbative and QMC calculations.

Figure~\ref{ee} shows the entanglement entropies of the ground state (triangle), hole-side excited state (square) and particle-side excited state (circle) calculated using perturbation theory. 
We find that the entanglement entropy of the hole-side excited state is much larger than the entanglement entropy of the particle-side excited state. This observation supports our intuition that the shift of the lobe boundary is closely related to the inter-species entanglement, with a larger shift on the hole-side of the boundary being accompanied by a larger inter-species entanglement compared to the particle-side of the boundary. As the hopping is increased we observe a decrease in the entanglement entropy of the hole-side excited state. This is due to the proliferation of quantum fluctuations which results in a finite contribution of Fock states with particle-hole excitations to the hole-excited state. As a consequence, a decrease in mutual information is observed as more sites corresponding to holes of component-a are available to be occupied by particle-b in each Fock state, with maximal mutual information occurring in the limit of zero hopping (and a single hole) as discussed in Section~\ref{sec3}.
The entanglement entropy of the particle-excited state, instead, shows a minimum as the hopping amplitude is increased. This could be explained by 
a two-fold effect of quantum fluctuations. On the one hand, the finite contribution of Fock states with particle-hole excitations tends to increase the mutual information due to the presence of holes, absent in the zero hopping limit. On the other hand, an initial proliferation of these states with very low weight might result in a decrease of entanglement entropy due to non-linear nature of the latter. This topic is under investigation at the moment~\cite{inprogress1}.
Lastly, Figure~\ref{ee} shows that the entanglement entropy of the ground state does not change significantly as $T^a/U^a$ is increased and it remains basically the same as  its value at zero hopping, i.e. $\mathbf{e}(\rho)=0$, corresponding to absence of inter-species entanglement.

\section{Conclusions } 
\label{sec8}

In summary, we have studied the inter-species entanglement of bosonic mixtures trapped in 
optical lattices within a perturbative approach.
Motivated by the observation that, in the presence of a second component, the Mott-insulator 
lobe shifts {\em{differently}} on the hole- and particle-side with respect to the Mott lobe 
of the single species system, we have investigated how this effect is related to the 
inter-species entanglement. This relationship indicates that inter-species entanglement 
plays an important role in
the characterization
of the quantum-phase transitions of mixtures, specifically in the Mott-insulator to superfluid 
transition. 

Our perturbative calculation is formulated in a Hilbert space decomposed by means of lattice symmetries (graph automorphisms). 
Within this decomposition, we have shown that if the ground state is {\em{not}} invariant under 
the independent action of symmetry operations on the two species, then the ground state must be 
entangled. 
The decomposition of the Hilbert space also results in a drastic reduction of the dimension of 
the Hilbert space {\em{relevant}} to the calculations of interest and hence a drastic reduction of the 
numerical cost of these calculations. 
We have calculated the Mott-lobe boundaries in the presence of a single particle of the second 
component and shown that, in the limit of small hopping, the hole-side of the boundary is 
dramatically affected by the presence of this single particle. 
We have compared our results with quantum Monte Carlo simulations by the two-worm algorithm. 
We have then quantified the entanglement in the Mott insulator ground state and the excited states
corresponding to the hole- and particle-side by calculating the entanglement entropy (von Neumann 
entropy) as the standard measure of bipartite entanglement. 
We have found that the entanglement entropy of the hole-side excited state is much larger than the 
entanglement entropy of the particle-side excited state. This means that the shift of the lobe 
boundary is closely related to the inter-species entanglement with a larger shift on the hole-side 
of the boundary being accompanied by a larger inter-species entanglement compared to the 
particle-side of the boundary.

A natural extension of the results presented in this paper is the study of the 
``structural'' nature of the entanglement as resulting from the ground state $|\Psi \rangle$ 
being written in terms of a decomposed Hilbert space~\cite{inprogress1}. 
Furthermore, this study can give us a better understanding of the dependence of entanglement 
entropy on the model parameters and better clarify the observations of Figure~\ref{ee}. Finally, 
we will study inter-species entanglement for Bose-Bose, Bose-Fermi, and Fermi-Fermi mixtures.

\section{Aknowledgments}
This work was supported by the NSF (PIF-1552978) and by MIUR (PRIN 2010LLKJBX). 
The computation for this project was performed at the OU Supercomputing Center 
for Education and Research (OSCER) at the University of Oklahoma (OU).


\appendix

\section{\label{app2}}

Assume $A_{g_1} \otimes B_{g_2} |\psi \rangle= |\psi \rangle$ for any $g_1, g_2\in G$. 
Then for two arbitrary $|{\bf n}, {\bf m} \rangle$ and $|{\bf n}', {\bf m}' \rangle$ in the 
product-configuration 
$(i,j)$ with $|{\bf n}', {\bf m}' \rangle =A_{g_1} |{\bf n} \rangle \otimes B_{g_2} |{\bf m} \rangle$, 
one has 
$$
\langle {\bf n}', {\bf m}' |\psi \rangle
=
\Bigl ( \langle {\bf n}| A^+_{g_1} \otimes \langle  {\bf m} | B^+_{g_2} \Bigr ) |\psi \rangle
=
\langle {\bf n},   {\bf m} | (A^+_{g_1} \otimes B^+_{g_2}) |\psi \rangle
= \langle {\bf n}, {\bf m} |\psi \rangle
\; .
$$
Therefore, $| \psi \rangle$ expands equally inside any product-configuration. 
By applying the trick used in Equation~\ref{Eq1}, we conclude 
$| \psi \rangle \in\bigoplus_{i,j}(\mathfrak{h}^a_i\bigotimes\mathfrak{h}^b_j)$. 
The reverse case is easily proved in two steps. First, we show that $\mathfrak{h}^a_i$ and 
$\mathfrak{h}^b_j$ are invariant under the action of $G$ using similar arguments as 
in proving $\mathfrak{h}^{ab}_k$ is invariant under the action of $G$. 
Then, we show that $\mathfrak{h}^a_i\otimes\mathfrak{h}^b_j$ is invariant under the 
action of $G \ast G$ using the definition of the representation 
$A\otimes B$. 
Thus, $\bigoplus_{i,j}(\mathfrak{h}^a_i\bigotimes\mathfrak{h}^b_j)$ is automatically 
invariant under the action of $G \ast G$.

%
\section{\label{app1}}
In this Appendix we want to show that a non-entangled $| \Psi \rangle$ belongs to 
$\bigoplus_{i,j}(\mathfrak{h}^a_i\bigotimes\mathfrak{h}^b_j)$. For this purpose, it is 
sufficient to show that, given 
$| {\bf n}'' ,  {\bf m}'' \rangle 
= A_{g} | {\bf n}\rangle \otimes  B_{g'} | {\bf m}\rangle$
one has,
$$
\langle  {\bf n}'' ,  {\bf m}''  | \Psi \rangle
= \langle {\bf n}| \otimes \langle  {\bf m} | \Bigl ( A^+_{g} \otimes  B^+_{g'} \Bigr )
| \Psi \rangle
=\langle {\bf n},  {\bf m}| \Psi \rangle 
$$
for arbitrary $g, g'\in G$ and arbitrary 
$| {\bf n},  {\bf m}\rangle = | {\bf n} \rangle \otimes | {\bf m}\rangle$. 
Let $| \Psi \rangle= | \phi^a \rangle \otimes | \phi^b \rangle$. Since $| \Psi \rangle$ 
is invariant under the action of $G$, we have
\begin{equation}
\label{A0}
\Bigl ( \langle  {\bf n} | A^+_{g}  \otimes  \langle {\bf m}| B^+_{g'} \Bigr ) | \Psi \rangle
=
\langle  {\bf n} | A^+_{g}  | \phi^a \rangle \; \langle {\bf m}| B^+_{g'} |\phi^b \rangle
%
=
\langle  {\bf n} | \phi^a \rangle \; \langle {\bf m}| \phi^b \rangle
\; .
\end{equation} 
Considering that $ \langle  {\bf n} | \phi^a \rangle\ne 0$ ($| \Psi \rangle$ has positive 
expansion coefficients), we have 
\begin{equation}
\label{A1}
\langle  {\bf m} | \phi^b \rangle
=\frac{ \langle  {\bf n} | A^+_{g}  | \phi^a \rangle }{ \langle  {\bf n} | \phi^a \rangle }
\langle  {\bf m} | B^+_{g'}  | \phi^b \rangle\; . 
\end{equation}
We can now rewrite Equation~\ref{A0} by replacing $| {\bf m} \rangle$ with 
$B_g | {\bf m} \rangle$, so that equation \ref{A1} becomes
%
%
%
\begin{equation}
\label{A2}
\langle  {\bf m} | B^+_{g} | \phi^b \rangle
=
\frac{ \langle  {\bf n} | A^+_{g}  | \phi^a \rangle }{ \langle  {\bf n} | \phi^a \rangle }
\langle  {\bf m} | (B^+_{g})^2  | \phi^b \rangle\; . 
%
\end{equation}
Then, inserting Equation~\ref{A2} in Equation~\ref{A1}, we have
\begin{equation}
\langle  {\bf m} | \phi^b \rangle
= \left ( \frac{ \langle  {\bf n} | A^+_{g}  | \phi^a \rangle }{ \langle  {\bf n} | \phi^a \rangle }
\right )^2
\langle  {\bf m} | (B^+_{g})^2  | \phi^b \rangle
\; . 
\end{equation}
We perform $n-1$ iterations of these steps, where $n$ is chosen to be the smallest 
integer such that $g^n=\mathbf 1$, and get
\begin{equation}
\langle  {\bf m} | \phi^b \rangle
= \left ( \frac{ \langle  {\bf n} | A^+_{g}  | \phi^a \rangle }{ \langle  {\bf n} | \phi^a \rangle }
\right )^n
\langle  {\bf m} | (B^+_{g})^n  | \phi^b \rangle
\end{equation}
Then, $ (B_{g})^n $ is the identity operator and thus
\begin{equation}
\label{A6}
\langle  {\bf m} | \phi^b \rangle
= \left ( \frac{ \langle  {\bf n} | A^+_{g}  | \phi^a \rangle }{ \langle  {\bf n} | \phi^a \rangle }
\right )^n
\langle  {\bf m} | \phi^b \rangle\; . 
\end{equation}
Since $\langle  {\bf m} | \phi^b \rangle \ne 0$ ($|\Psi \rangle$ has positive expansion coefficients), 
Equation~\ref{A6} implies 
$|\langle  {\bf n} | A^+_{g}  | \phi^a \rangle | =|\langle  {\bf n}| \phi^a \rangle |$. 
In a similar way, we can show 
$|\langle  {\bf m} | B^+_{g'}  | \phi^b \rangle|
= | \langle  {\bf m}| \phi^b \rangle | $. 

Finally, since $\langle {\bf n}| A^+_{g}|\phi^a \rangle \langle {\bf m}| B^+_{g'} |\phi^b \rangle>0$
and
$\langle  {\bf n} |\phi^a \rangle \langle  {\bf m} |\phi^b \rangle>0$, we have
$\langle {\bf n}| A^+_{g}|\phi^a \rangle \langle {\bf m}|B^+_{g'}|\phi^b \rangle
=
\langle  {\bf n} |\phi^a \rangle \langle  {\bf m} |\phi^b \rangle$
or
\begin{equation}
\langle {\bf n}'' ,{\bf m}''|\Psi \rangle
=
\langle  {\bf n}, {\bf m} |\Psi \rangle \; .
\end{equation}

\bigskip
\bigskip
\bigskip

\bibliography{Dec062015}

\end{document}